\documentclass[12pt,preprint]{aastex}

\def\ltsima{$\; \buildrel < \over \sim \;$}
\def\gtsima{$\; \buildrel > \over \sim \;$}
\def\lsim{\lower.5ex\hbox{\ltsima}}
\def\gsim{\lower.5ex\hbox{\gtsima}}
\def\lapp{\ifmmode\stackrel{<}{_{\sim}}\else$\stackrel{<}{_{\sim}}$\fi}
\def\gapp{\ifmmode\stackrel{>}{_{\sim}}\else$\stackrel{<}{_{\sim}}$\fi}

\def\kms{~{\rm km~ s}^{-1}}
\def\sigp{\sigma_{\rm P}}

\newdimen\minuswidth    %define @ width of minus sign for tables
\setbox0=\hbox{$-$}
\minuswidth=\wd0
\catcode`@=\active
\def@{\kern\minuswidth}
\setbox0=\hbox{\rm0}
\shorttitle{Resolved-star kinematics of NGC 6388}
\shortauthors{Lanzoni et al.}
 
\begin{document} 
\title{The velocity dispersion profile of NGC 6388 from resolved-star spectroscopy: no
  evidence of a central cusp and new constraints on the black hole mass\footnote{Based on
    observations collected at the European Southern Observatory, Cerro
    Paranal, Chile (under proposals 381.D-0329, 073.D-0211 and
    073.D-0760). Also based on observations with the NASA/ESA HST
    (Prop. 19835), obtained at the Space Telescope Science Institute,
    which is operated by AURA, Inc., under NASA contract NAS5-26555.}}

\author{
B. Lanzoni\altaffilmark{2}, 
A. Mucciarelli\altaffilmark{2}, 
L. Origlia\altaffilmark{3},
M. Bellazzini\altaffilmark{3},
F.R. Ferraro\altaffilmark{2},
E. Valenti\altaffilmark{4},
P. Miocchi\altaffilmark{2}, 
E. Dalessandro\altaffilmark{2},
C. Pallanca\altaffilmark{2},
D. Massari\altaffilmark{2}
}
\affil{\altaffilmark{2} Dipartimento di Fisica e Astronomia, Universit\`a degli
  Studi di Bologna, Viale Berti Pichat 6/2, I--40127 Bologna, Italy}
\affil{\altaffilmark{3} INAF- Osservatorio Astronomico di Bologna, Via
  Ranzani, 1, 40127 Bologna, Italy}
\affil{\altaffilmark{4} European Southern Observatory,
  Karl-Schwarzschild-Strasse 2, 85748 Garching bei M\"unchen, Germany}
\date{9 April, 2013}

\begin{abstract}
By combining high spatial resolution and wide-field spectroscopy
performed, respectively, with SINFONI and FLAMES at the ESO/VLT we
measured the radial velocities of more than 600 stars in the direction
of NGC 6388, a Galactic globular cluster which is suspected to host an
intermediate-mass black hole.  Approximately 55\% of the observed
targets turned out to be cluster members.  The cluster velocity
dispersion has been derived from the radial velocity of individual
stars: 52 measurements in the innermost $2\arcsec$, and 276 stars
located between $18\arcsec$ and $600\arcsec$.  The velocity dispersion
profile shows a central value of $\sim 13\kms$, a flat behavior out to
$\sim 60\arcsec$ and a decreasing trend outwards.  The comparison with
spherical and isotropic models shows that the observed density and
velocity dispersion profiles are inconsistent with the presence of a
central black hole more massive than $\sim 2000 M_\odot$. These
findings are at odds with recent results obtained from integrated
light spectra, showing a velocity dispersion profile with a steep
central cusp of $23-25\kms$ at $r<2\arcsec$ and suggesting the
presence of a black hole with a mass of $\sim 1.7 \times 10^4 M_\odot$
\citep{lutz11}.
We also found some evidence of systemic rotation with amplitude
$A_{\rm rot}\sim 8 \kms$ in the innermost $2\arcsec$ (0.13 pc),
decreasing to $A_{\rm rot}=3.2 \kms$ at $18\arcsec<r<160\arcsec$.
\end{abstract}


\section{INTRODUCTION}
\label{sec:intro}
The confirmation of the existence of intermediate mass ($10^3-10^4
M_\odot$) black holes (IMBHs) would have a dramatic impact on a number
of open astrophysical problems, ranging from the formation of
supermassive BHs and their co-evolution with galaxies, to the origin
of ultraluminous X-ray sources in nearby galaxies, up to the detection
of gravitational waves \citep[e.g.,][]{gebh05}.  However, the evidence
gathered so far in support of the existence of IMBHs are inconclusive
and controversial \citep[see, e.g.,][]{noyola10, jay10}. Galactic
globular clusters (GCs) are thought to be the best places where to
search for these elusive objects. In fact, the extrapolation of the
"Magorrian relation" \citep{magorrian} down to the IMBH masses
naturally leads to the GC mass regime. Moreover, numerical simulations
have shown that the cores of GCs are the ideal habitat for the
formation of IMBHs \citep[e.g.][]{portegies04}.  For these reasons,
the recent years have seen an increasing number of works dedicated to
the search for IMBHs in GCs, exploiting all observational channels,
ranging from the detection of X-ray and radio emission \citep[see][and
  references therein]{strader12, kirsten12}, to the detailed study of
the cluster structure, especially in terms of the shape of the density
and velocity dispersion profiles \citep[e.g.,][hereafter L11]{gebh00,
  gerssen02, lanz07_imbh, noyola10, jay10, lutz11}.  In spite of such
an effort, however, no firm conclusions could be drawn to date.  This
is mainly because of the great difficulties encountered from both the
theoretical and the observational points of view.  In particular, the
expected X-ray and radio emission are still quite uncertain. A
power-law density profile $\Sigma_*(r)\propto r^\alpha$ with a slope
$\alpha \sim -0.3$ (significantly shallower than the one expected in a
post-core collapse system: $\alpha\sim -0.7$) is expected in the
innermost cluster regions \citep[e.g.][]{baumg05,miocchi07}, even
though this could be a non-univocal signature of a central IMBH
\citep{vesperini10}.  In the presence of an IMBH a steep central cusp
is expected also in the velocity dispersion profile
\citep{baumg05,miocchi07}. However, the precise determination of the
stellar density and (especially) the velocity dispersion is very
challenging in the highly crowded central regions of GCs.  In
particular, measuring the projected velocity dispersion ($\sigp$) in
GCs has always been a very difficult task. In fact, GCs are close
enough for allowing to resolve single stars; hence, the estimate of
$\sigp$ from the line broadening of integrated light spectra can be
easily falsified by the dominant contribution of a few bright objects
\citep[e.g.,][]{dubath97}. On the other hand their stellar densities
(especially in the centre) are so high that the measurement of
\emph{individual radial velocities} is very challenging from the
observational point of view and it requires very high spatial
resolution spectroscopy. For the accurate determination of the
internal proper motions, both high spatial resolution imaging and long
enough time baselines are needed
\citep[e.g.][]{mcnamara03,jay10,mcnamara12}.  Finally, different
methodologies for measuring the velocity dispersion could bring to
incompatible results \citep[see the case of $\omega$ Centauri
  discussed in][]{vdm10,noyola10}.

This paper is devoted to NGC\,6388. Using high spatial resolution
($0.025\arcsec\times 0.027\arcsec$/pix) data acquired with the HST/ACS
High Resolution Channel (HRC), we discovered a shallow central cusp
($\alpha=-0.2$) in the inner $1\arcsec$ and, from the comparison with
King models including a central IMBH \citep{miocchi07}, we suggested
that a $\sim 6\times 10^3 M_\odot$ BH could be hidden in the center of
this cluster \citep{lanz07_imbh}. However, subsequent investigations,
both in the X-ray and in the radio bands, argued against such a
conclusion \citep{nucita08, cseh10, bozzo11}.  The question has been
recently re-opened by L11, who used \emph{integrated light spectra}
acquired with the FLAMES/ARGUS to derive $\sigma_P$ from the spectral
line broadening: they found an increasing trend toward the cluster
centre, up to values of $23-25 \kms$, and concluded that this is best
fitted by assuming the presence of an IMBH of $(1.7\pm 0.9)\times 10^4
M_\odot$.  Here we derive the velocity dispersion profile of NGC 6388
from the \emph{radial velocities of more than 300 individual
  stars}. We find no evidence of central cusp in the velocity
dispersion and the comparison with theoretical models suggests that
this is inconsistent with the presence of an IMBH more massive than
$\sim 2000 M_\odot$.

The paper is organized as follows. Observations and data analysis are
described in Section \ref{sec:obs}.  Results are presented in
Sect. \ref{sec:resu} and include the determination of the cluster
membership (Sect. \ref{sec:sele}), the study of its systemic rotation
(Sect. \ref{sec:rot}) and the measurement of the velocity dispersion
profile (Sect. \ref{sec:vdisp}). Section \ref{sec:discuss} is devoted
to the comparison with previous results and theoretical models.  The
conclusions are presented and discussed in Sect. \ref{sec:concl}.
Appendix A is dedicated to the discussion of the impact of unresolved
background and stellar blends on the determination of radial
velocities.

\section{OBSERVATIONS AND DATA ANALYSIS}
\label{sec:obs}
\subsection{The SINFONI data set for the cluster central region}
\label{sec:sinfo}
In order to acquire spectra of individual stars in the center of NGC
6388 we exploited the high spatial resolution capabilities of SINFONI
\citep{eisen}, a near-IR (1.1-2.45 $\mu$m) integral field spectrograph
fed by an adaptive optics module and mounted on the YEPUN (VLT-UT4)
telescope at the ESO Paranal Observatory.  By using the 100 mas
plate-scale and the $K$-band grating (sampling the 1.95-2.45 $\mu$m
wavelength range), we derived spectra at a resolution $R=4000$ for
$\sim 60$ stars located in a $3.2\arcsec\times 3.2\arcsec$ region
centered on the cluster gravity center \citep[as quoted
  in][]{lanz07_imbh}.  The observations have been performed in service
mode between April and June 2008 (ESO proposal ID: 381.D-0329(A), PI:
Lanzoni), under an average seeing of $\sim 0.8\arcsec$ (FWHM). A star
of magnitude $R\sim 12$ located $\sim 9\arcsec$ from the cluster
center has been used as natural guide star for adaptive optics
correction, leading to an average Strehl ratio\footnote{The Strehl
  ratio one of the main parameters characterizing the image quality of
  data taken with adaptive optics systems and it corresponds to the
  amount of light contained in the diffraction-limited core of the
  point spread function, with respect to the total flux.} of
$\sim30\%$.  The high level of crowding combined with the small field
of view prevented us from estimating the sky background through a
dithering technique. Hence, for background subtraction purposes the
observations were performed with a target-sky-sky-target sequence,
using an empty sky region located $\sim 2\arcmin$ North-East from the
cluster center. A single target frame is the combination of 10
exposures, each one 20 seconds long. The observations have been
repeated until a total exposure time of $\sim 2.2$ hr on target was
reached.

\begin{figure}[!hp]
\begin{center}
\includegraphics[scale=0.7]{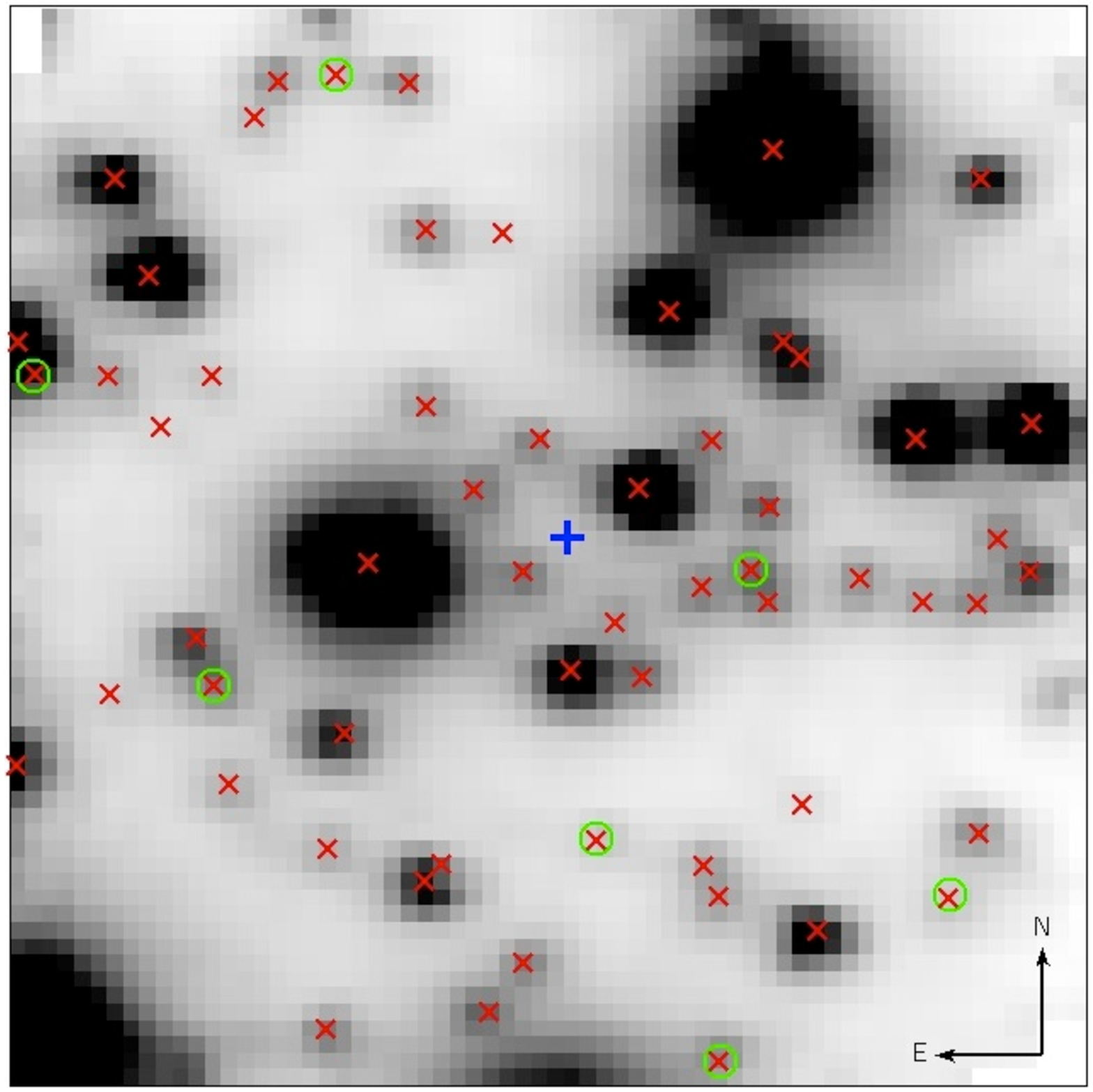}
\caption{Reconstructed SINFONI image. The field of view is
  $3.2\arcsec\times 3.2\arcsec$, North is up, East is left. Red
  crosses mark the spaxels from which we extracted the spectra,
  corresponding to stellar centroids measured in the HST/ACS-HRC
  image. Green circles mark the four stars for which the radial
  velocity could not be measured and the three targets conservatively
  rejected because significantly blended with a close brighter star
  (see Sect. \ref{sec:vr} and Appendix A). The large blue cross flags
  the adopted position of the cluster gravity centre
  \citep[from][]{lanz07_imbh}.}
\label{sinfo_ima}
\end{center}
\end{figure}

\begin{figure}[!hp]
\begin{center}
\includegraphics[scale=0.8]{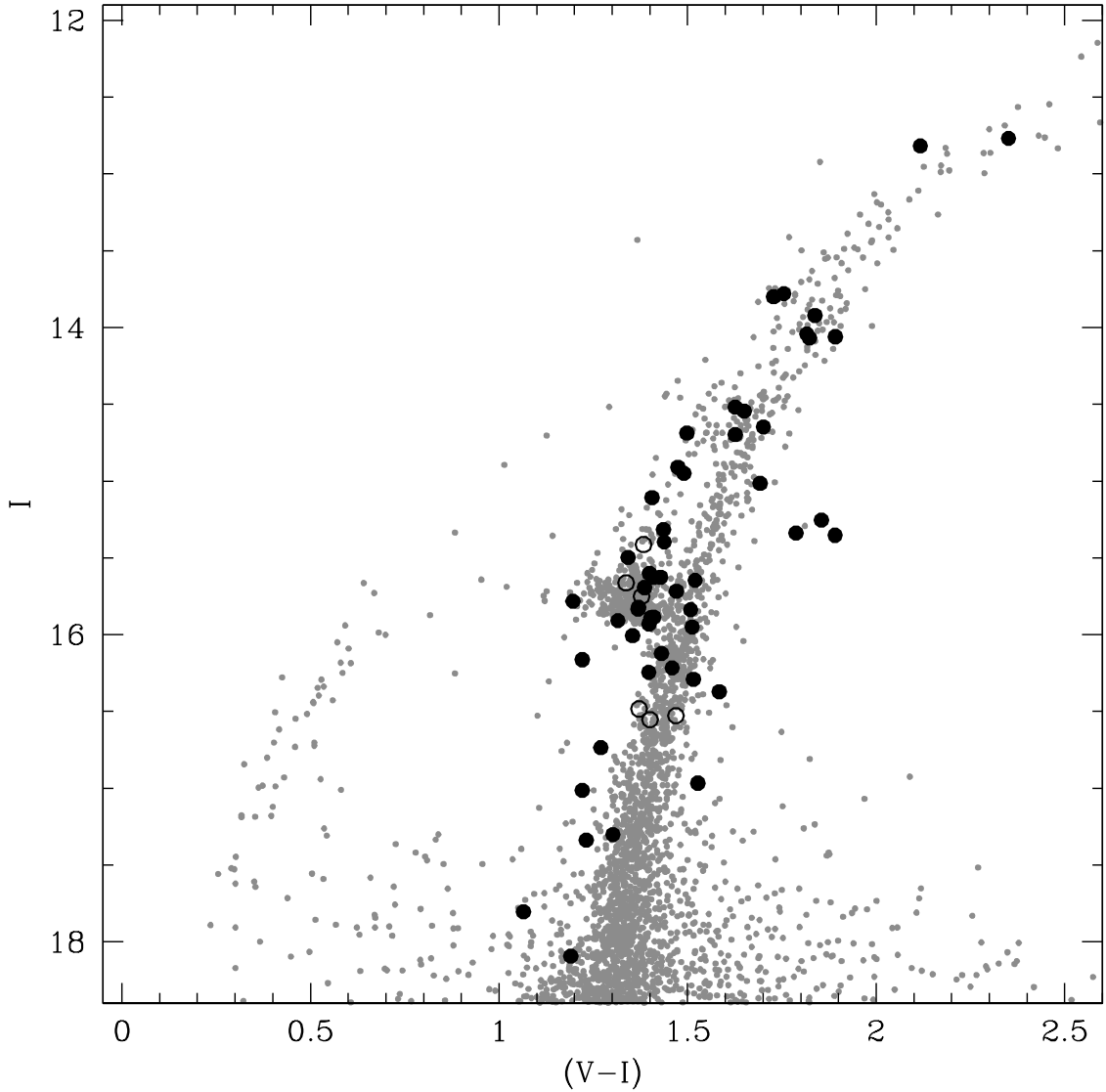}
\caption{HST/ACS-HRC color-magnitude diagram of the central region of
  NGC 6388.  The 59 stars for which we extracted a SINFONI spectrum
  are highlighted as large circles: solid circles correspond to the 52
  objects for which we could reliably measure the radial velocity,
  empty circles flag the remaining seven stars.}
\label{sinfo_cmd}
\end{center}
\end{figure}

We derived a wavelength calibrated 3D data cube, as well as a
reconstructed 2D image, by using the SINFONI pipeline v.2.3.2
\citep{modigliani10}. Following the pipeline prescriptions, we first
corrected all the target and sky data for darks, flats, geometrical
distortions and differential atmospheric refraction. We then
subtracted the sky background by using the observations of the sky
field, we calibrated in wavelength by using a Th-Ar reference arc
lamp, and we finally combined the corrected target frames to build the
3D data cube.  In principle this provides a total of $32\times 64$
spectra.  In practice, in order to minimize the contamination of the
spectrum from the light of neighbouring stars, we made the
conservative choice of extracting only the spectrum of the central
(and most exposed) spaxel corresponding to the star centroid.  To this
end we performed the photometric analysis of the reconstructed SINFONI
image by using ROMAFOT \citep{romafot} and thus determined the
position of the source centroids.  Then, the HST/ACS-HRC catalogue
discussed in \citet{lanz07_imbh} was roto-translated on the SINFONI
one. The resulting astrometric solution has an accuracy of less than
0.2 spaxels and provided us with a clear-cut identification of 90\% of
the SINFONI targets. In the remaining cases (corresponding to a
candidate SINFONI target surrounded by a few close, fainter
companions), a careful visual inspection of the SINFONI image was
necessary to precisely locate the spaxel corresponding to the brighter
star.  With IRAF we thus extracted the 1D spectrum for a total of 59
spaxels (Figure \ref{sinfo_ima}) corresponding to 59 stars in the
magnitudes range $13\lsim I\lsim 18$ (Figure \ref{sinfo_cmd}). The
extracted spectra have been cleaned from the telluric absorption lines
through the division by a spectrum of a featureless early-type star
observed each night with the same instrument setup of the science
target.  The typical signal-to-noise (S/N) ratio \emph{per pixel} at
the wavelength $2.18 \mu$m ranges from $\sim 20$ at $I= 18$, to $\sim
80$ for the brightest sources.

\subsection{The FLAMES data set for the cluster external regions}
\label{sec:flames}
The spectra of individual stars beyond the cluster core have been
acquired by using the ESO VLT multi-object spectrograph FLAMES
\citep{pasquini02}, in the MEDUSA UVES+GIRAFFE combined mode. This
provides 132 fibers to observe an equivalent number of targets in one
single exposure, over a field of view of $25\arcmin$ in diameter.
Under proposal 381.D-0329(B) (PI: Lanzoni) four pointings of 2320 s
each have been performed in June and July 2008 with the GIRAFFE
grating HR21 (which samples the Ca II triplet spectral range at a
resolution $R=16200$) and the UVES setup Red Arm 580, covering the
wavelength range $4800 \mathring{\rm A}<\lambda<6800\mathring{\rm A}$
at a resolution $R=47000$.  The targets have been selected from the
combined HST/ACS-WFC and ESO/WFI photometric catalog already discussed
in \citet{lanz07_imbh} and \citet{ema08}, considering only isolated
objects (with no brighter stars within a circle of $1\arcsec$ radius),
having $V<17$ and being located along the canonical evolutionary
sequences of the color-magnitude diagram (CMD).  About 15 GIRAFFE
fibers and 2 UVES fibers in each pointing have been used to sample the
sky. A total of 422 and 7 stars were observed with GIRAFFE and UVES,
respectively.  Additional MEDUSA data have been retrieved from the ESO
Archive.  These include the spectra of red giant branch stars observed
with GIRAFFE gratings HR11 and HR13 under program 073.D-0211 (PI:
Carretta), with HR13 under program 073.D-0760 (PI: Catelan), and with
UVES Red Arm 580 setup in both cases. While some of these stars are in
common with the sample discussed above, 99 objects are new. The FLAMES
data set therefore consists of a total of 528 individual star spectra.

All these data have been reduced with the most updated versions of the
UVES and GIRAFFE ESO pipelines,
%\footnote{http://www.eso.org/sci/software/pipelines/},.........
including bias subtraction, flat-field correction, wavelength
calibration through a reference Th-Ar lamp, rebinning to a constant
pixel-step, extraction of the one dimensional spectra and (for the
UVES targets only) order merging.  The accuracy of the wavelength
calibration was checked by measuring several emission sky lines in the
available spectral ranges and comparing the observed wavelengths to
the rest-frame positions quoted by \citet{oster}.  The removal of the
sky background is particularly crucial for the observations performed
with the GIRAFFE setup HR21, since the third component
($\lambda=8662\mathring{\rm A}$) of the Ca II triplet is heavily
affected by O$_2$ emission lines. For targets observed with HR11/HR13
and Red Arm 580 setups, instead, the sky level is only a few percent
of the star one (because of the brightness of the targets),
introducing only a negligible amount of noise in the stellar
spectra. In all cases the sky background has been removed by
subtracting from each stellar spectrum a master sky spectrum obtained
as the median of the observed sky spectra. Each individual sky
spectrum has been inspected by eye in order to check for any possible
contamination from close stellar components and a few sky spectra
showing absorption lines due to the light of close stars have been
discarded.  Note that the employed pre-reduction procedure takes into
account the fiber to fiber relative transmission, in order to properly
scale the intensity of each spectrum to the different fiber response.
For the targets observed with HR11 and/or HR13 setups, multiple
exposures of the same stars are available. In these cases we added
together the sky-subtracted spectra to increase the S/N ratio.  The
same procedure has been adopted for the UVES targets with multiple
exposures.  The final S/N per pixel is always larger than 50.

\subsection{Stellar radial velocities}
\label{sec:vr}
\begin{figure}[!hp]
\begin{center}
\includegraphics[scale=0.8]{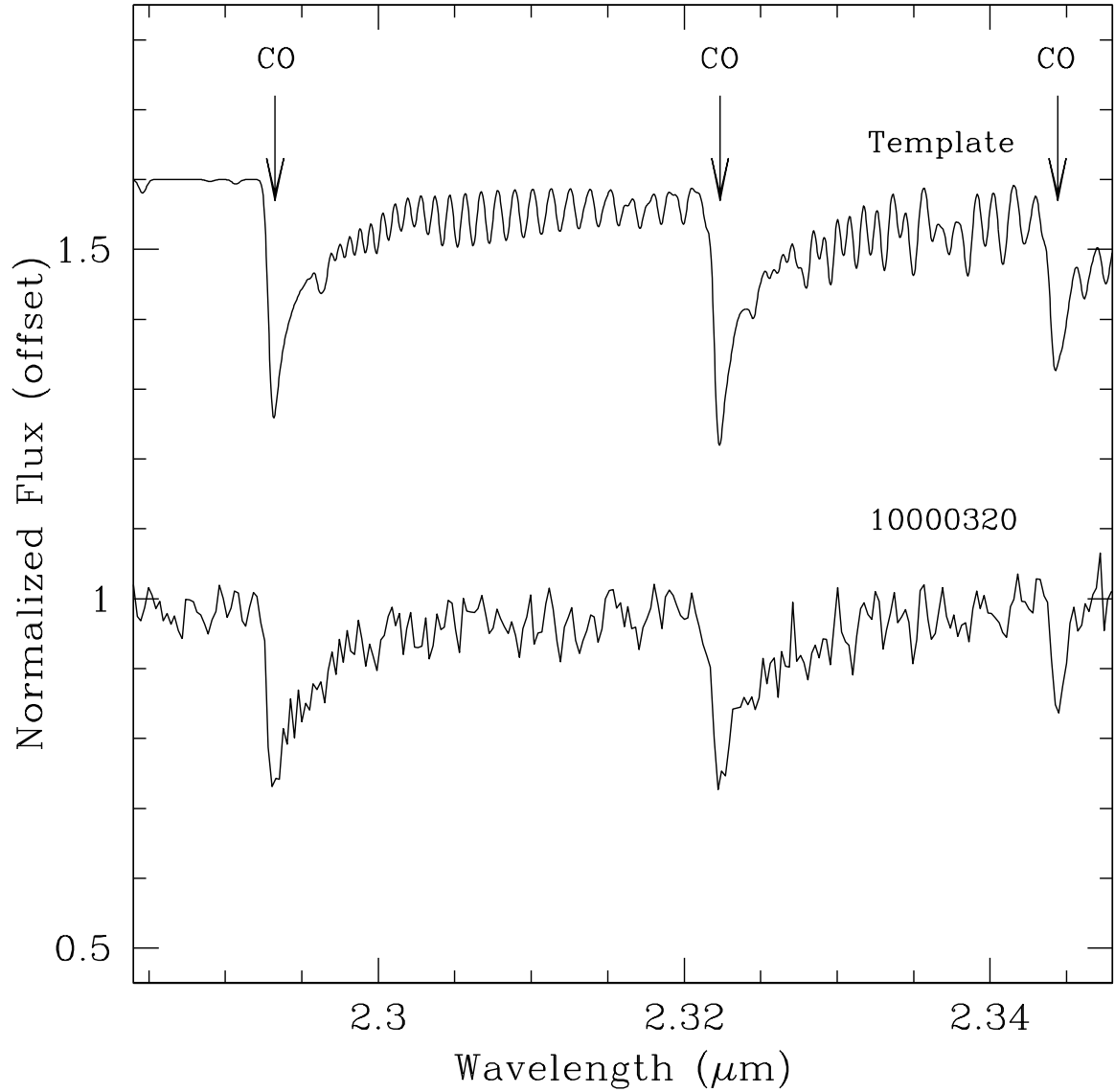}
\caption{Comparison between one spectrum acquired with SINFONI and the
  appropriate template. Three CO band-heads used to compute the
  stellar radial velocities are flagged.}
\label{sinfo_spec}
\end{center}
\end{figure}

The radial velocities ($V_r$) of the SINFONI targets have been
measured from the $^{12}$C$^{16}$O band-heads (see Figure
\ref{sinfo_spec}) and, occasionally, a few atomic lines. For the
FLAMES targets we used several atomic lines, depending on the GIRAFFE
setup. In all cases we adopted the Fourier cross-correlation method
\citep{tonry79} as implemented in the {\sl fxcor} IRAF task.  The
observed spectrum is cross-correlated with a template of known radial
velocity and a cross-correlation function (i.e., the probability of
correlation as a function of the pixel shift) is computed.  Then, this
is fitted by using a Gaussian profile and its peak value is
derived. Once the spectra are wavelength calibrated, the pixel shift
obtained by the Gaussian fit is converted into radial velocity.  We
employed different reference template spectra according to the adopted
spectral configuration and stellar type.  Basically, we used suitable
synthetic spectra computed in the wavelength range covered by the
instrumental setup and convolved with a Gaussian profile to reproduce
the spectral resolution of the used gratings.  For the SINFONI targets
we used synthetic spectra computed with the code described by
\citet{origlia93} and following the same procedure (i.e., model
atmospheres, atomic and molecular line list) described in
\citet{origlia04}.  For the FLAMES targets we computed synthetic
spectra with the Linux version of the SYNTHE code \citep{sbordone05}.

Four SINFONI spectra are of low quality. Three of them are among the
faintest stars in the $K$-band and their spectra have a very low S/N
ratio ($<5$), probably because of a bad response of the corresponding
spaxels.  The spectrum of the star located at the south edge of the
field is unusable as a consequence of the adopted jittering
procedure. Hence they have been excluded from the subsequent
analysis. After a careful analysis of the effects of unresolved
background and blended sources on the determination of the radial
velocity (see Appendix A), three additional targets have been removed
and the final SINFONI sample consists of 52 stars located between
$0.2\arcsec$ and $1.9\arcsec$ from the cluster center. From the FLAMES
spectra we measured the radial velocity of all the analyzed targets,
corresponding to a total of 528 stars located at distances ranging
between $11\arcsec$ and $883\arcsec$.

The velocity errors are computed for each star following the
prescriptions by \citet{tonry79}, which take into account the accuracy
in the fit of the peak of the correlation function, the errors due to
mismatches between object and template spectra and the effect of the
anti-symmetric noise component of the correlation function. The
typical uncertainties are of a few $\kms$ for the SINFONI targets,
$0.4-0.5\kms$ for the FLAMES giants, and $1.5-2\kms$ for the FLAMES
horizontal branch stars because of the blending between the Ca II
lines and the Hydrogen Paschen lines.  All radial velocities have been
reported in the same heliocentric reference system, by applying the
corresponding correction computed with the IRAF task {\sl rvcorrect}.
The presence of several targets in common among the different FLAMES
datasets allowed to verify possible off-sets in the radial velocity
zero point, due to the different spectral configurations and
wavelength calibrations.  We assumed as reference sample the one
observed with the HR21 setup, since it represents more than 70\% of
the entire dataset.  No relevant offset is found when this is compared
with the other samples, the average radial velocity differences being
smaller than $1 \kms$, with a standard deviation $<1.5\kms$. This
guarantees a good internal homogeneity of our radial velocity zero
point and confirms the reliability of our error estimates.

\subsection{Stellar metallicities}
\label{sec:met}
With the aim of providing an additional constraint to the cluster
membership, we also computed the metallicity of our target stars.
Unfortunately, a precise measure for the SINFONI targets is prevented
by the low spectral resolution of the available data. We could just
derive a rough estimate of the target metallicities from a few Fe, Ca
and Na lines, suggesting that, within a large uncertainty ($\sim 0.2$
dex), they are compatible with the cluster mean value
\citep[{[Fe/H]}$=-0.44\pm 0.01$; ][]{carretta07}. To measure the
metallicity of the FLAMES targets different methodologies have been
adopted according to the different spectral configurations used. The
atmospheric parameters (effective temperature and gravity) have been
estimated for each target from the comparison between the star
position in the CMD and theoretical isochrones by \citet{pietrinf06}.
For the stars observed with gratings HR11 and HR13 and with UVES, we
performed a direct estimate of the metallicity from the observed
equivalent widths (EWs) of the Fe I lines. Given the target
atmospherical parameters and the instrumental spectral resolution, we
selected only the Fe I lines predicted to be unblended. EWs were
measured by means of DAOSPEC \citep{stetson}. Details about the choice
of the atomic data and the computation of model atmospheres and
chemical abundances are in \citet{muccia12}.  In the case of targets
observed with the HR21 grating, with the exception of very cold stars
and blue horizontal branch stars (for which the Ca II lines are
heavily affected by strong TiO molecular bands and Hydrogen Paschen
lines, respectively), the metallicity has been derived from the Ca II
lines by following the approach described in \citet{carrera07}.  The
metallicity has been measured for a total of 508 stars.

\section{RESULTS}
\label{sec:resu}
\subsection{Cluster membership}
\label{sec:sele}
\begin{figure}[!hp]
\begin{center}
\includegraphics[scale=0.8]{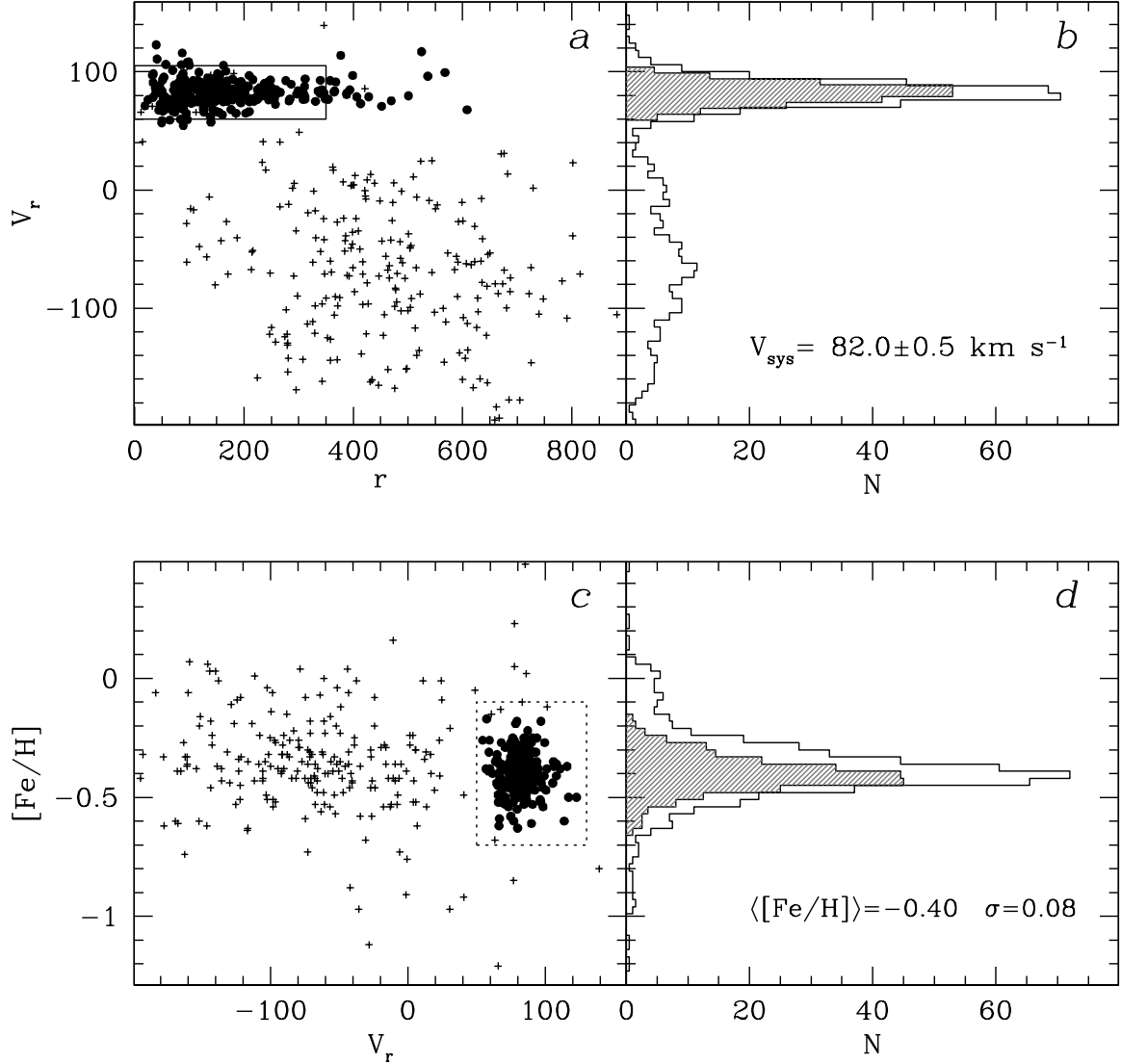}
\caption{Radial velocity and metallicity distributions (top and bottom
  panels, respectively) of the FLAMES targets used to determine the
  cluster systemic velocity, mean metallicity and membership.  The
  mean metallicity of NGC 6388 (see label in Panel $d$) has been
  estimated from the (280) stars enclosed in the dotted box
  ($50<V_r<130\kms$ and $-0.7<$[Fe/H]$<-0.1$) drawn in Panel $c$. Only
  the (276) stars having a metallicity value within $3\sigma$ from the
  mean have been considered as cluster members (solid circles in Panel
  $c$). Their metallicity distribution is compared to that of the
  entire population in Panel $d$ (shaded and empty histograms,
  respectively). The radial velocities of the 276 cluster members
  (solid circles) are shown as a function of the distance from the
  cluster centre in Panel $a$. The corresponding number distribution
  is plotted as shaded histogram and compared to that of the entire
  FLAMES sample (empty histogram) in Panel $b$.  The box drawn in
  Panel $a$ encloses the sub-sample of members used to conservatively
  compute the cluster systemic radial velocity, which is labelled in
  Panel $b$.  }
\label{vrfe}
\end{center}
\end{figure}

The radial velocities measured from the FLAMES spectra are plotted as
a function of the distance from the cluster center ($r$) in the
top-left panel of Figure \ref{vrfe}.  Most of the observed stars
clearly clump around $V_r\sim 80\kms$, while the remaining ones show a
large spread at lower, mostly negative values. Hence, a selection in
$V_r$ has been used as driving criterion to distinguish cluster
members from field objects.

\begin{figure}[!hp]
\begin{center}
\includegraphics[scale=0.8]{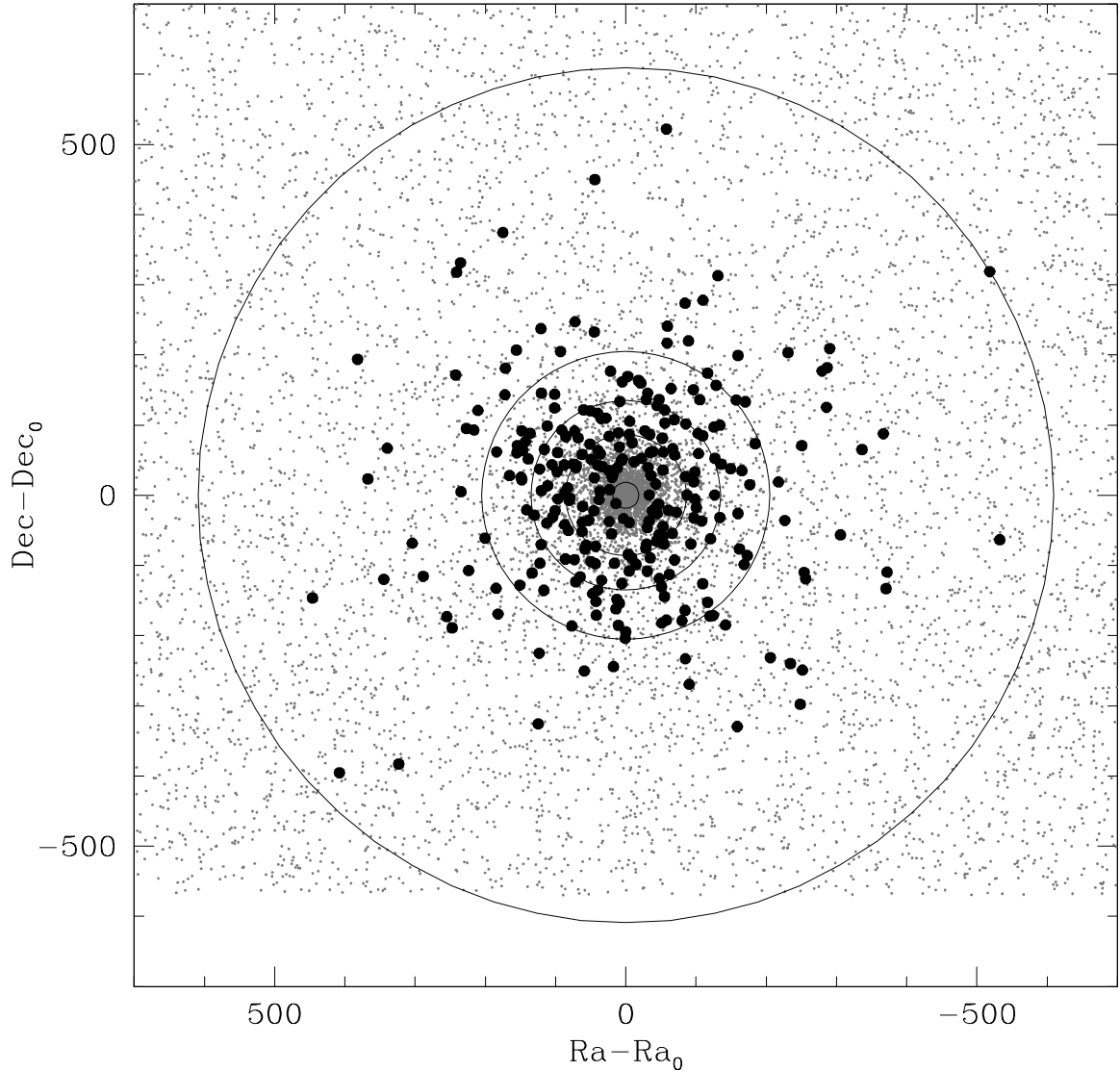}
\caption{Map of stars brighter than $V=18$, derived from the combined
  HST/ACS-WFC and ESO/WFI photometric catalog discussed in
  \citet{ema08}.  The 276 FLAMES targets selected as cluster members
  are highlighted as large solid circles.  The concentric annuli used
  to compute the cluster velocity dispersion profile in the external
  regions (see Table \ref{tab_vdisp}) are also marked.}
\label{flames_map}
\end{center}
\end{figure}

\begin{figure}[!hp]
\begin{center}
\includegraphics[scale=0.8]{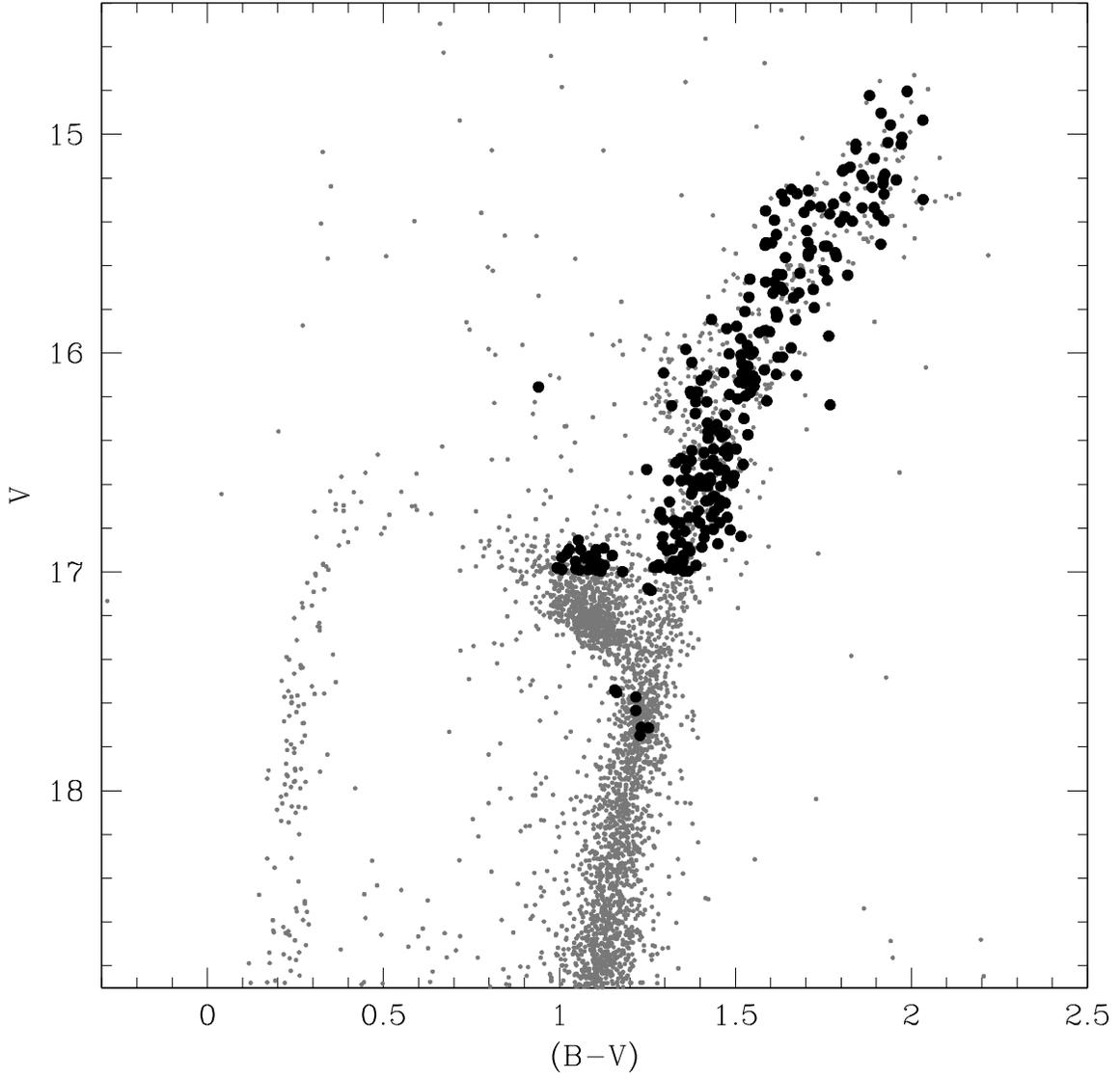}
\caption{($V, B-V$) color-magnitude diagram of NGC 6388 obtained from
  HST/ACS-WFC and ESO-WFI observations \citep{ema08}.  The 276 FLAMES
  targets selected as cluster members are highlighted as large solid
  circles.}
\label{flames_cmd}
\end{center}
\end{figure}

As apparent from the bottom panels of Fig. \ref{vrfe}, the separation
between cluster members and field stars is less clear in terms of
metallicity.  This is indeed expected, since the field population is
constituted by disk giants with [Fe/H] ranging between zero and $-1.0$
\citep{besancon}.  Hence, to determine the average metallicity of NGC
6388 we selected (280) stars with $50<V_r<130\kms$; in addition, based
on the star distribution in the ([Fe/H], $V_r$) plane, we excluded the
most evident outliers by applying a preliminary cut in metallicity:
$-0.7<$[Fe/H]$<-0.1$ (see the dotted box in the bottom-left panel of
Fig. \ref{vrfe}). The average value turns out to be
$\langle$[Fe/H]$\rangle =-0.40$ with a dispersion of 0.08 dex, in good
agreement with \citet{carretta07}.  We then conservatively considered
as cluster members only those stars having a metallicity value within
$3\sigma$ from the mean:\footnote{Note that by estimating the
  atmospheric parameters from a theoretical isochrone appropriate for
  NGC 6388 (see Sect.  \ref{sec:met}) we implicitly assumed that all
  stars belong to the cluster. While this is not necessarily true, the
  Besancon Galactic Model \citep{besancon} shows that the expected
  contamination from field stars in the color, magnitude and radial
  velocity ranges covered by our targets is quite low ($\sim
  3.5\%$). Moreover, adopting an isochrone appropriate for the field
  contaminating population ([Fe/H]=0, distance =8 kpc), the resulting
  metallicities turn out to be only a few 0.1 dex larger. In any case,
  we stress that the driving selection criterion for cluster
  membership is the radial velocity, while metallicity is used just to
  refine the selection, even at the cost of rejecting a few cluster
  members. Most importantly, we stress that any metallicity selection
  is unable to introduce kinematical biases in the analysis.}  this
yields to 276 stars (54\% of the total) with $-0.64<$[Fe/H]$<-0.16$,
which are marked as solid circles in the bottom- and top-left panels
of Fig. \ref{vrfe}. Their metallicity distribution is compared to that
of the entire population (508 objects) in the bottom-right panel of
the figure.  No significant changes are found for reasonable
modifications of the assumed cuts in radial velocity and
metallicity. Moreover, no significant impact is expected on the
following analysis by the potential inclusion of a few interlopers.
The resulting 276 FLAMES cluster members are highlighted as solid
circles in Figures \ref{flames_map} and \ref{flames_cmd} which show,
respectively, the cluster map and CMD obtained from the photometric
catalog presented in \citet{ema08}.  The radial velocities of SINFONI
targets range between 56 and $104\kms$. By also taking into account
their very central radial position ($r<2\arcsec$), we consider all of
them as cluster members.

To determine the systemic velocity ($V_{\rm sys}$) of NGC 6388 we
adopted even more conservative limits: out of the 276 cluster members,
only (240) stars with $r<350\arcsec$ and $60\le V_r\le 105\kms$ were
selected (see the solid box in the top-left panel of Fig. \ref{vrfe}).
The mean value, computed with a Maximum-Likelihood method, turns out
to be $V_{\rm sys}=82.0\pm 0.5 \kms$.  Different, but still
reasonable, assumptions about the limits in cluster-centric distance
and/or radial velocity cuts do not produce any significant change in
this result, which is also in agreement with the systemic velocity
quoted in the literature for NGC 6388 \citep[see, e.g., Harris 1996 --
  2010 version;][]{casetti10}.  The coordinates, magnitudes and radial
velocities (referred to $V_{\rm sys}$) measured for all (52+276)
cluster members are listed in Table \ref{tab_vr} (where we also give
the metallicities of FLAMES targets). In the rest of the paper,
$\widetilde V_r$ will indicate radial velocities referred to the
cluster systemic velocity: $\widetilde V_r\equiv V_r-V_{\rm sys}$.

\subsection{Systemic rotation}
\label{sec:rot}
\begin{figure}[!hp]
\begin{center}
\includegraphics[scale=0.8]{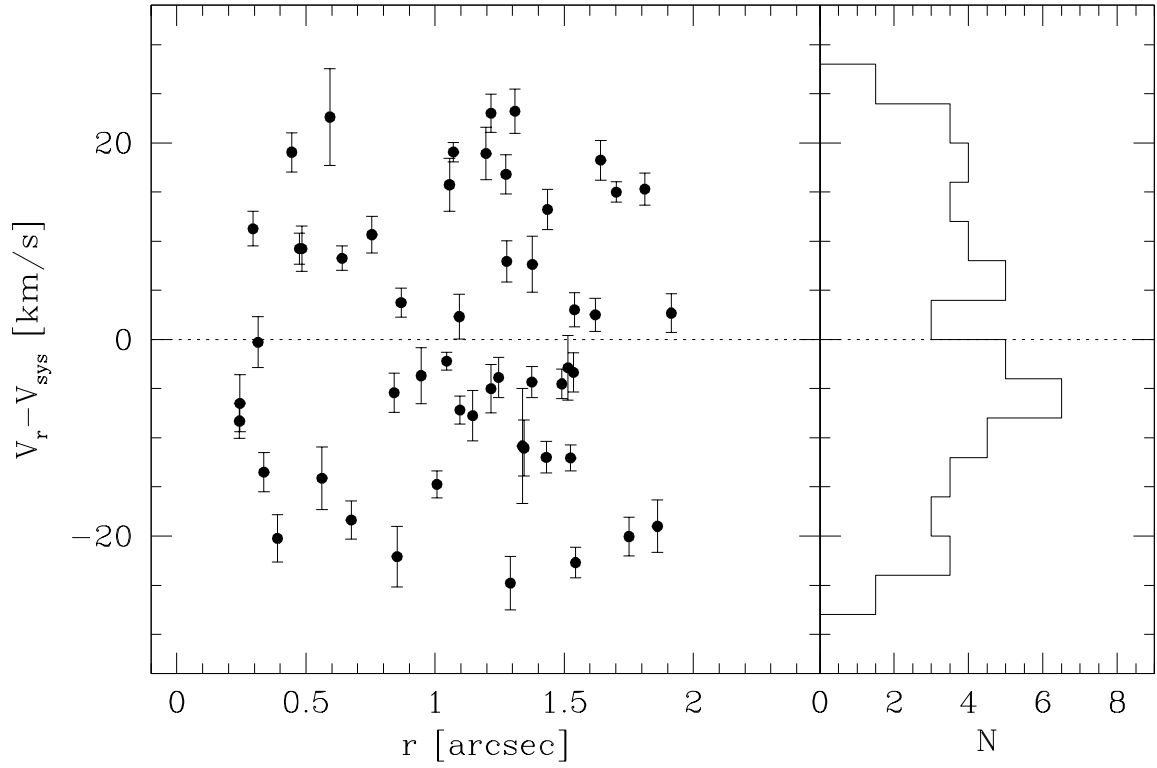}
\caption{{\it Left panel:} Distribution of $V_r-V_{\rm sys}$ for the
  52 SINFONI targets. The histogram in the right-hand panel shows a
  bimodality indicating the presence of systemic rotation.}
\label{vrr_sinfo}
\end{center}
\end{figure}

\begin{figure}[!hp]
\begin{center}
\includegraphics[scale=0.8]{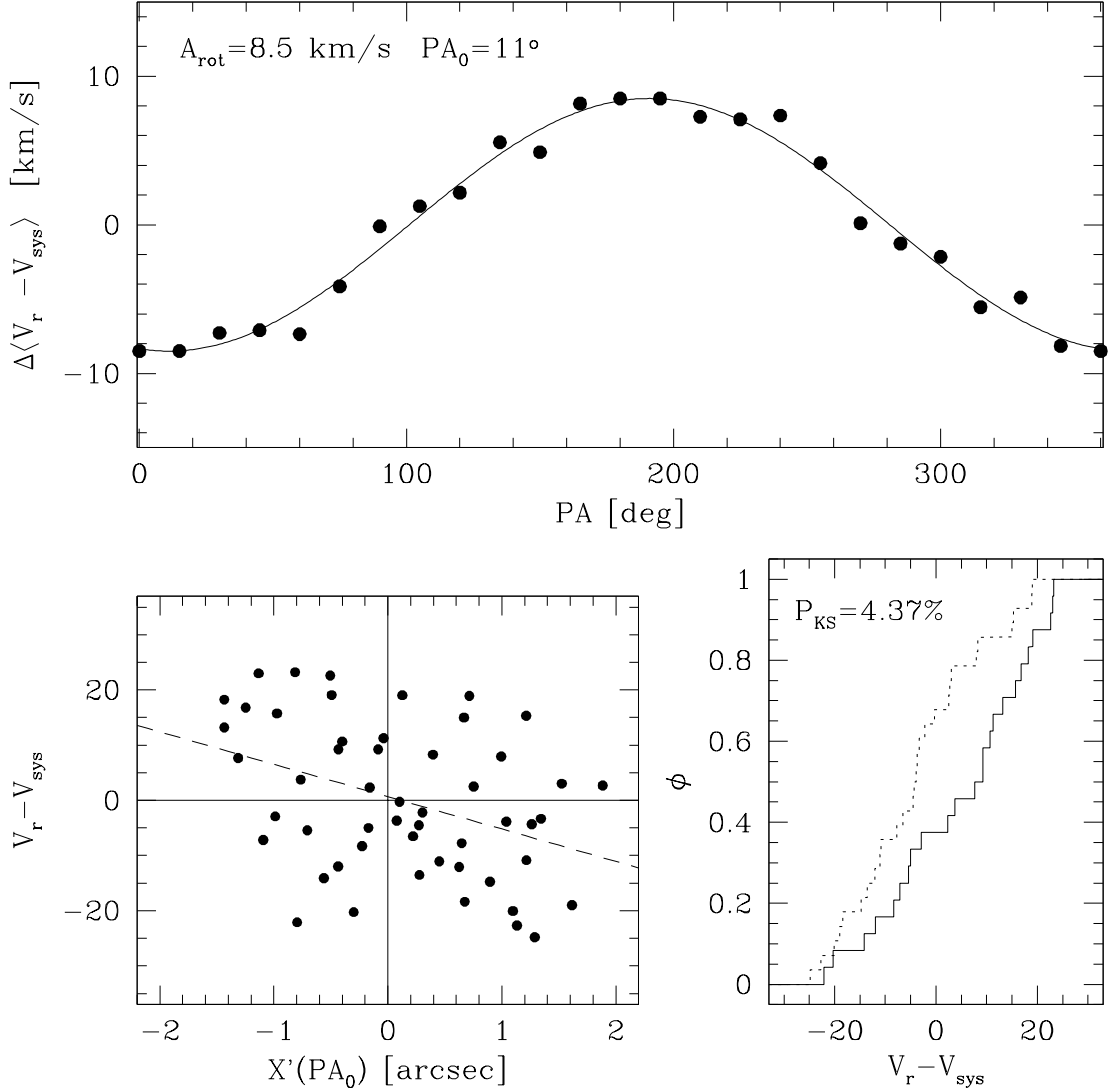}
\caption{{\it Upper panel:} Difference between the mean radial
  velocities of the SINFONI targets located on each side of the
  cluster with respect to a line passing through the center and having
  position angle PA (measured from North, PA=$0\arcdeg$, to East,
  PA=$90\arcdeg$), as a function of PA. The solid line is the sine
  function that best fits the observed pattern. The corresponding
  values of the rotation amplitude ($A_{\rm rot}$) and the position
  angle of the rotation axis (PA$_0$) are labelled. {\it Lower left
    panel:} Distribution of $V_r-V_{\rm sys}$ as a function of the
  distance $X'$ from the center projected onto the axis (dashed line)
  perpendicular to the best-fit rotation axis. {\it Lower right
    panel:} Cumulative radial distribution of stars having $X'({\rm
    PA}_0)>0$ (solid line) and $X'({\rm PA}_0)<0$ (dotted
  line). P$_{\rm KS}$ quotes the Kolmogorov-Smirnov probability that
  the two distributions are drawn from the same parent population.}
\label{vrot_sinfo}
\end{center}
\end{figure}

Figure \ref{vrr_sinfo} shows $\widetilde V_r$ as a function of the
distance from the cluster center for the 52 SINFONI targets. A
bimodality is apparent in this distribution and suggests the presence
of systemic rotation. To investigate this possibility, we used the
method fully described in \citet[][and references
  therein]{bellaz12}. We considered a line passing through the cluster
center with position angle PA varying between $0\arcdeg$ (North
direction) and $90\arcdeg$ (East direction), by steps of $15\arcdeg$.
For each value of PA, such a line splits the SINFONI sample in
two. The difference $\Delta\langle \widetilde V_r\rangle$ between the
mean radial velocity of the two sub-samples was computed and it is
plotted as function of PA in the upper panel of Figure
\ref{vrot_sinfo}.  Its coherent sinusoidal behavior is a signature of
rotation and the parameters of the best-fit sine function provide us
with the position angle of the rotation axis (PA$_0=11\arcdeg$) and
the amplitude of the rotation curve \citep[$A_{\rm rot}=8.5 \kms$; for
  the exact meaning of this parameter see the discussion
  in][]{bellaz12}. The corresponding rotation curve is shown in the
lower-left panel of Fig. \ref{vrot_sinfo}. The Kolmogorov-Smirnov
probability that the distributions of $\widetilde V_r$ for the two
sub-samples on each side of the rotation axis are drawn from the same
parent population is $\sim 4\%$ (lower-right panel), indicating a not
particularly high statistical significance, possibly due to the
limited sample.  We verified that taking into account or ignoring such
a rotation has only very weak effects on the computation of the
cluster velocity dispersion, which is dominant. While a rotation or
shearing signature within $\sim 3\arcsec$ from the centre is claimed
also by L11, it seems to be mainly caused by the shot noise produced
by three bright stars in the field of view (see their Figure 7).  In
any case, our best-fit position angle of the rotation axis seems not
to coincide with their.

\begin{figure}[!hp]
\begin{center}
\includegraphics[scale=0.8]{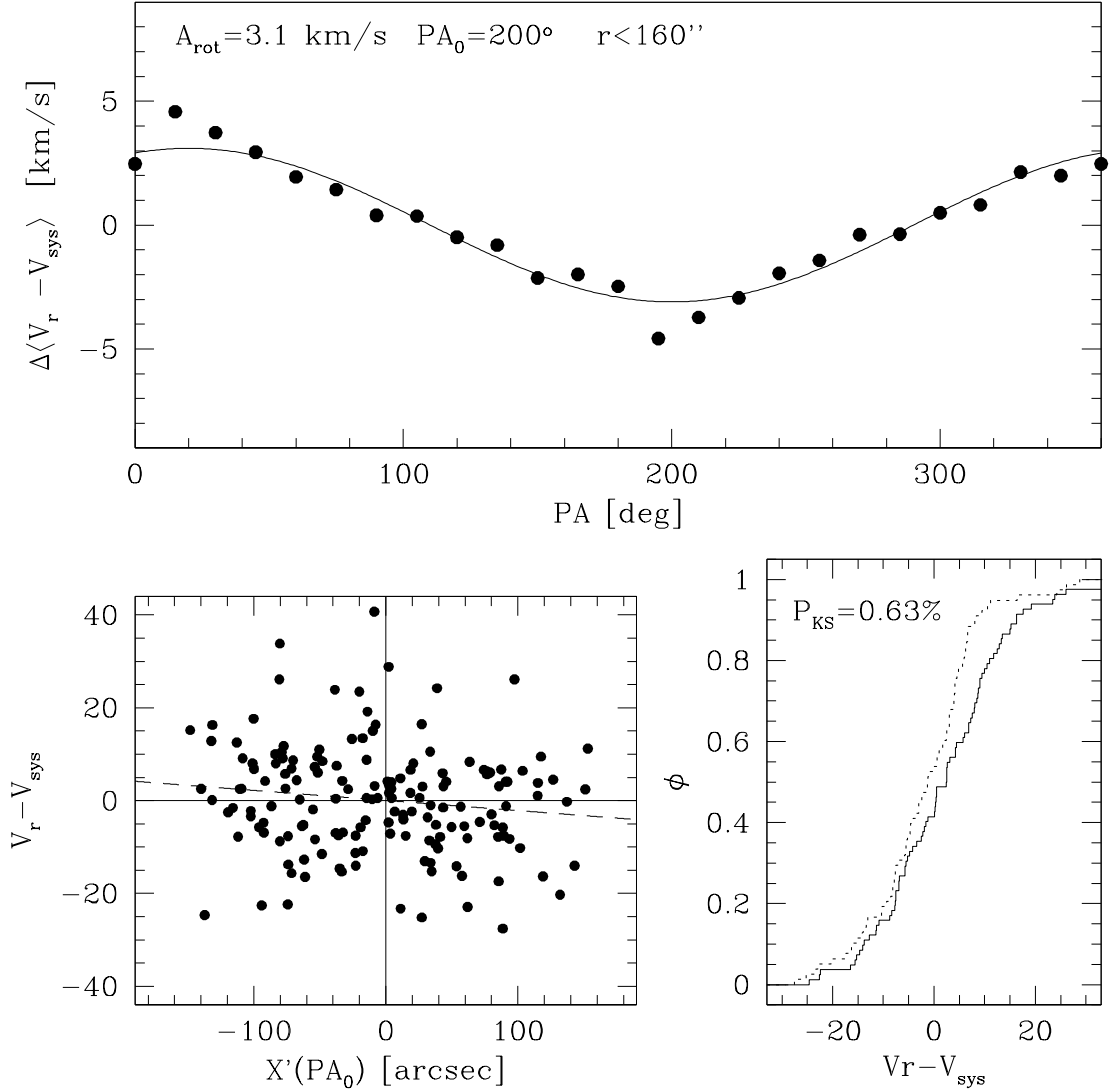}
\caption{As in Fig. \ref{vrot_sinfo}, but for the 160 FLAMES targets
  located between $18\arcsec$ and $160\arcsec$ from the cluster
  centre.}
\label{vrot_flames}
\end{center}
\end{figure}

We performed a similar analysis also for the FLAMES sample,
considering various distance intervals from the cluster centre.  The
strongest signal of systemic rotation is found for stars with
$18\arcsec<r<160\arcsec$ (160 in total). The corresponding parameters
are PA$_0=200\arcdeg$ and $A_{\rm rot}=3.2\kms$, and the
Kolmogorov-Smirnov probability for a common origin of the $\widetilde
V_r$ distributions for the two sub-samples is of only 0.6\% (see
Figure \ref{vrot_flames}). No evidence of significant rotation is
instead detected in the outermost cluster regions, where the second
moment of the velocity distribution decreases and it becomes
increasing difficult to measure independently the ordered and the
random motions.  These results do not confirm those recently published
by \citet{bellaz12}, who suggest that some rotation is present between
$\sim 160\arcsec$ and $400\arcsec$ from the center, based, however, on
a small sample of less than 40 stars (in the same radial range we
count, instead, more than 100 stars).

We finally emphasize that the results presented in L11 are all in
terms of the second moment of the velocity distribution ($V_{\rm RMS}
= \sqrt{V^2_{\rm rot} + \sigma^2_P}$, with $V_{\rm rot}$ indicating
the rotational velocity). Hence, a direct comparison with their work
(Sect. \ref{sec:cfr_L11}) requires to analyze the same quantity.  Also
for this reason, exactly as in L11, we do not apply any correction for
rotation and we deal with the second moment of the velocity
distribution. However, because the rotation is small, in the following
we continue to discuss in terms of velocity dispersion and we use
symbol $\sigma_P$ instead of $V_{\rm RMS}$ all over the paper.

\subsection{Velocity dispersion}
\label{sec:vdisp}
\begin{figure}[!hp]
\begin{center}
\includegraphics[scale=0.8]{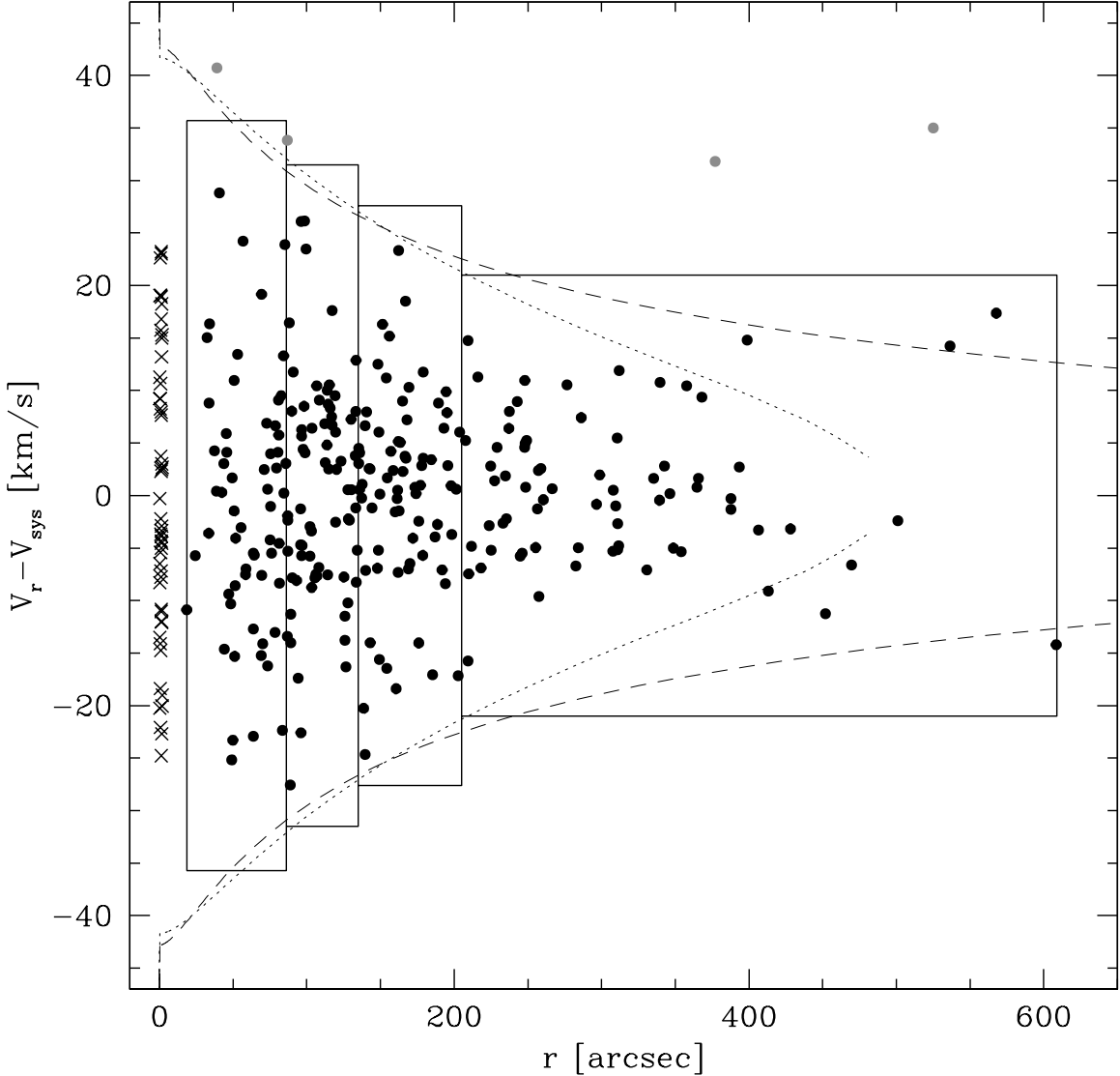}
\caption{Radial velocity (referred to the cluster systemic velocity)
  as a function of the distance from the center, for the selected
  cluster members: 52 SINFONI targets are shown as crosses, 276 FLAMES
  stars are plotted as solid circles. The radial bins (see Table
  \ref{tab_vdisp}) adopted to determine the external portion of the
  velocity dispersion profile from the FLAMES data are marked as
  rectangles. Their vertical size corresponds to the $3\sigma$ range
  of the local velocity dispersion $\sigp(r)$; stars not included in
  these boxes (and not used in the computation of $\sigp$) are marked
  as grey circles. The dotted and dashed curves mark the $\pm 3\sigma$
  velocity dispersion profiles of the best-fit King and Wilson models,
  respectively (see Sect. \ref{sec:cfr_mod}).}
\label{vrr}
\end{center}
\end{figure}

\begin{figure}[!hp]
\begin{center}
\includegraphics[scale=0.8]{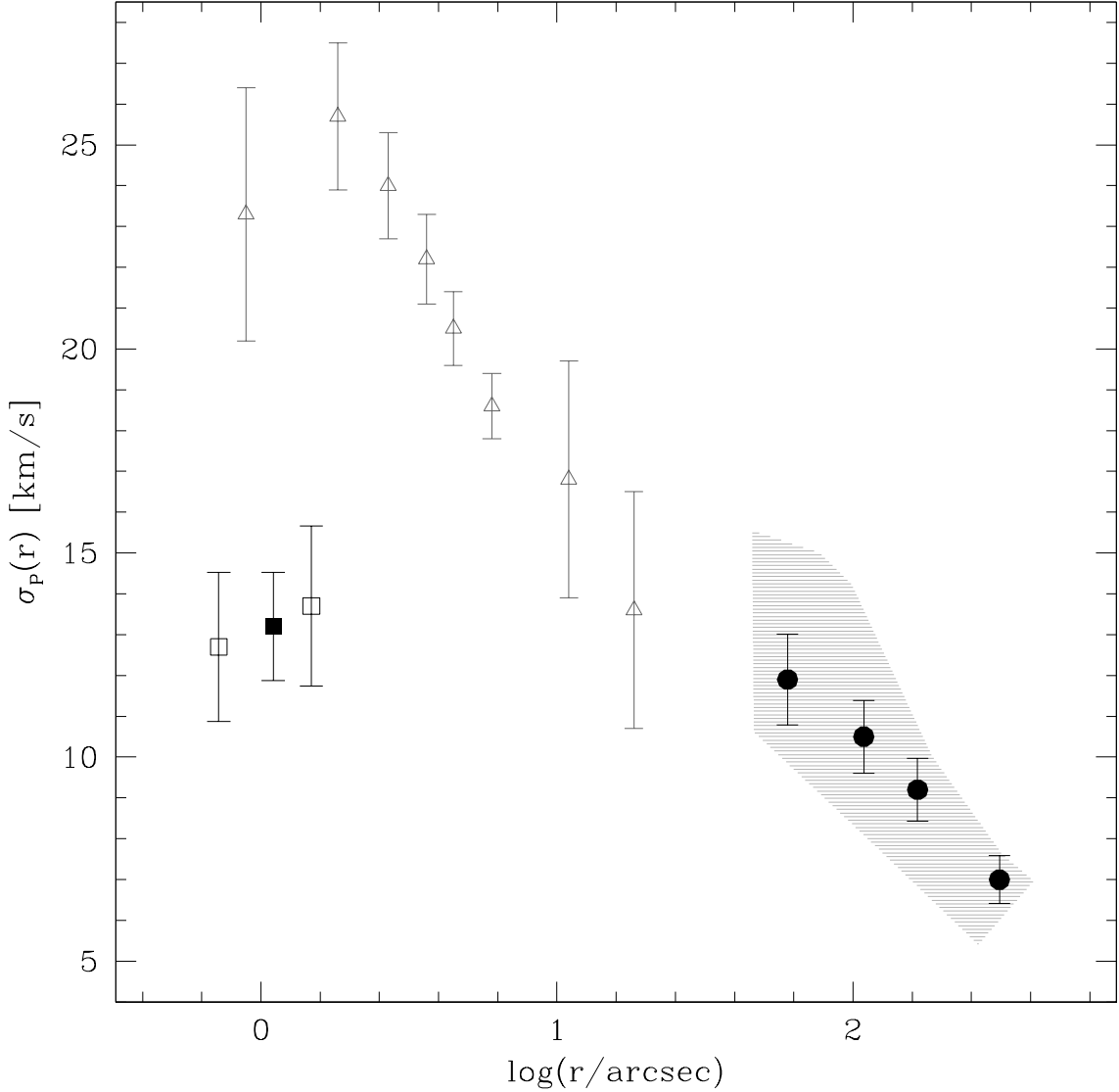}
\caption{Projected velocity dispersion profile of NGC 6388 as
  determined from the radial velocity of individual stars (squares and
  circles).  The solid square marks the value of $\sigp$ obtained from
  the entire SINFONI sample (52 stars at $r<1.9\arcsec$), the two
  empty squares correspond to the values derived within and beyond
  $r=1.2\arcsec$ (26 and 26 stars in the two bins,
  respectively). Results from the FLAMES dataset are shown as black
  circles, with the grey region indicating the dispersion obtained for
  different choices of the radial bins. Triangles correspond to the
  results obtained by L11 from integrated light spectroscopy and are
  shown for comparison.  }
\label{vdisp}
\end{center}
\end{figure}

The radial distribution of $\widetilde V_r$ for all the selected
cluster members is shown in Figure \ref{vrr}. To compute the projected
velocity dispersion profile $\sigp(r)$, the surveyed area has been
divided in a set of concentric annuli, chosen as a compromise between
a good radial sampling and a statistically significant number ($\gsim
50$) of stars.  In each radial bin, $\sigp$ has been computed from the
dispersion of the values of $\widetilde V_r$ measured for all the
stars in the annulus, by following the Maximum Likelihood method
described in \citet[][see also Martin et al. 2007; Sollima et
  al. 2009]{walker06}. An iterative $3\sigma$ clipping algorithm was
applied in each bin (the stars thus excluded from the computation are
marked as grey circles in Fig. \ref{vrr}).  The error estimate is
performed by following Pryor \& Meylan (1993).  The resulting velocity
dispersion profile is shown in Figure \ref{vdisp} and listed in Table
\ref{tab_vdisp}.  Given the number of stars in the SINFONI data set
(52 objects in total) we computed $\sigp(r)$ by considering both one
single central bin (solid square in the figure) and two separate
annuli (26 stars at $r<1.2\arcsec$, 26 stars beyond; see the empty
squares). For the external data set we tried different sets of radial
bins (grey region in Fig. \ref{vdisp}) and the values obtained by
adopting the four annuli listed in Table \ref{tab_vdisp} are shown as
black circles.  These results show that the velocity dispersion of NGC
6388 has a central ($r\sim 1\arcsec$) value of $\sim 13\kms$, stays
approximately flat out to $r\sim 60\arcsec$, and then decreases to
$\sim 7\kms$ at $200\arcsec<r <600\arcsec$.  If no $\sigma$ clipping
algorithm is applied, the velocity dispersion profile remains almost
unchanged, with the only exception of the outermost data-point that
rises to $\sim 8.9\pm0.8\kms$.

\section{DISCUSSION}
\label{sec:discuss}
\subsection{Comparison with previous studies}
\label{sec:cfr_L11}
The inner part ($r< 10\arcsec$) of the velocity dispersion profile
derived by L11 from the line broadening of integrated-light spectra
(triangles in Fig. \ref{vdisp}) is clearly incompatible with the one
obtained in this study from the radial velocities of individual stars
(squares).  While the exact reason for such a disagreement is not
completely clear, a detailed comparison between our radial velocity
measurements and L11 radial velocity map (which is available in
electronic form and shown in their Figure 7) suggests that the shot
noise corrections applied by these authors may have been insufficient.

\begin{figure}[h]
\begin{center}$
\begin{array}{cc}
\includegraphics[width=8.5cm]{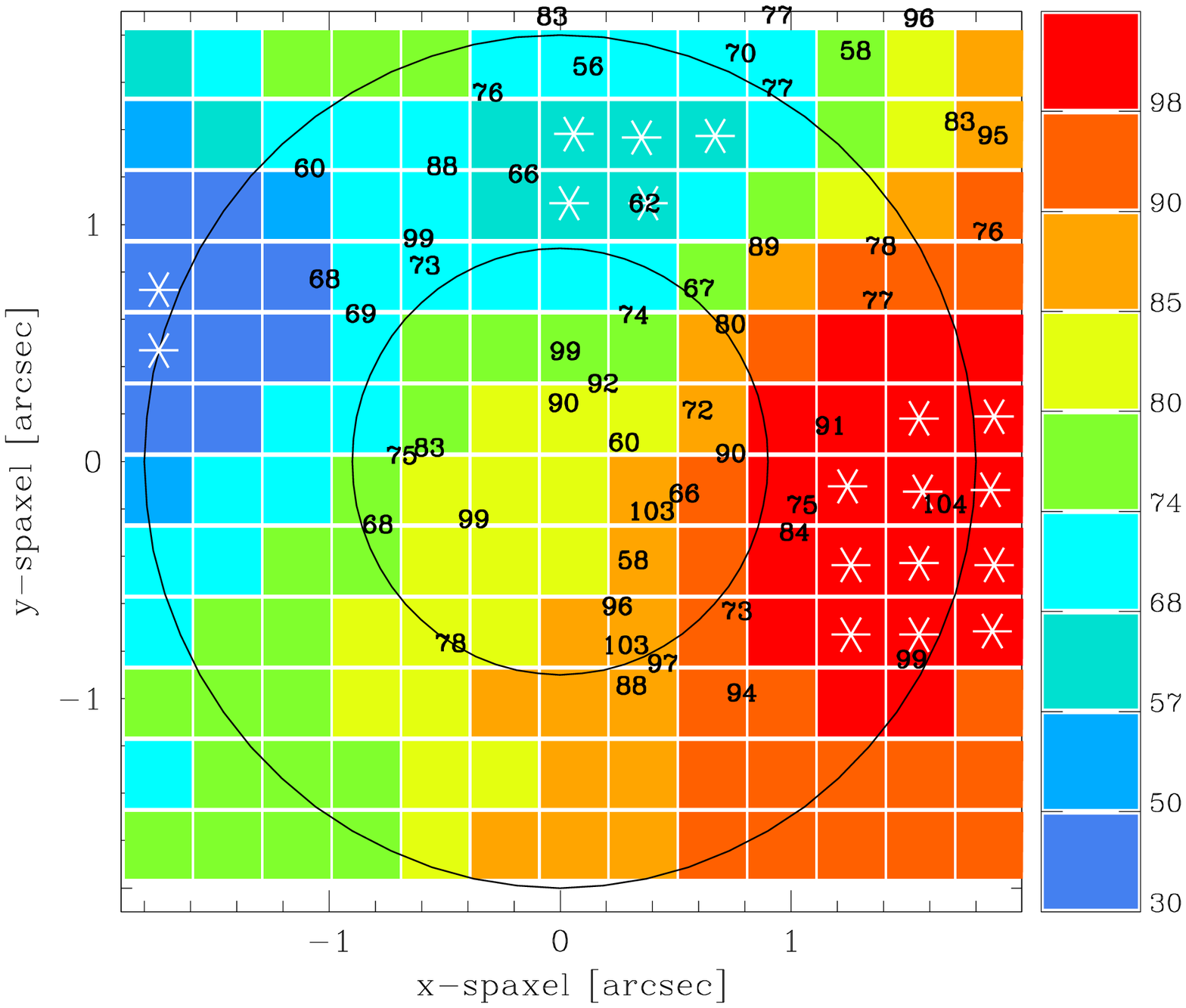} &
\includegraphics[width=6.5cm]{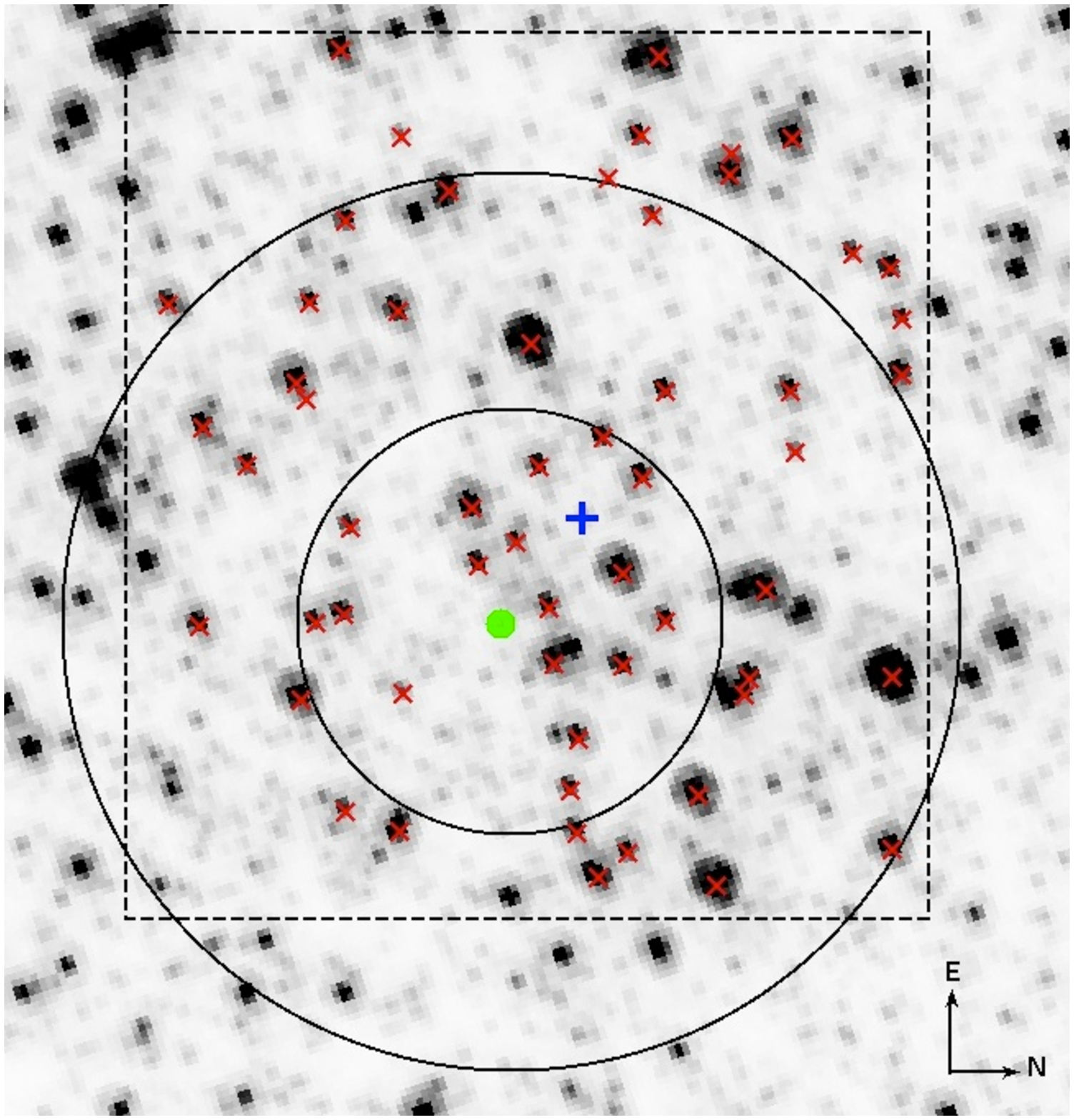}
\end{array}$
\end{center}
\caption{{\it Left panel:} Approximate reproduction of L11's Figure 7,
  showing in colors their radial velocity map for the innermost
  $2\arcsec\times 2\arcsec$ (North is right, East is up). The values
  of $V_r$ increase from $\sim 30\kms$ (dark blue), to more than
  $100\kms$ (red), as indicated in the color-bar on the right-hand
  side. White asterisks mark the spaxels that L11 excluded from the
  analysis to correct for the shot noise effect due to the presence of
  very bright stars. The values of the radial velocities that we
  measure for each individual star detected in the same field of view
  are marked in black at the star position. The two circles (of radius
  $0.9\arcsec$ and $1.9\arcsec$) correspond to the two innermost
  radial bins considered in L11. {\it Right panel:} $I$-band
  HST/ACS-HRC image of the same cluster region.  The orientation of
  the map and the two circles are as in the left-hand panel, while the
  dashed box indicates the field of view of our SINFONI observations.
  Red crosses mark the stars for which we measured the radial velocity
  $V_r$, which is labelled in black in the left panel of the
  figure. The blue cross marks the position of the gravity centre used
  in the present work \citep[from][]{lanz07_imbh}, while the green
  circle indicates the cluster centre adopted in L11.}
\label{vmap}
\end{figure}

Figure \ref{vmap} approximately reproduces the central $2\arcsec\times
2\arcsec$ region of L11's Figure 7, with the values of $V_r$ that we
measured for each individual star in the same field of view marked in
black.  In the second annulus, for instance, we measure radial
velocities as low as $73, 75, 77\kms$ in regions where they quote $V_r
>98\kms$ (see the red spaxels on the right-hand side). We also find
$V_r=88, 99\kms$ where they measure $V_r<74 \kms$ (cyan region on the
upper-left side). This might be because those spaxels in L11 are still
highly contaminated, respectively, by the very bright stars with $V_r>
100\kms$ and $V_r\sim 62 \kms$ (in both studies) that produce the
large red spot on the right-hand side of the figure and the dark-cyan
spot at the top, despite the applied shot noise correction (see the
white asterisks flagging a few spaxels in these red and dark cyan
regions).  More in general, we note that, even in the case of close
stars, our individual radial velocity values can show significant
differences, thus indicating that the measurements are independent
each other. Instead, the $V_r$ color codes of L11 are quite uniform in
several spaxels around the brightest stars (i.e., the degree of
correlation, or cross-talking, is high).  Even though L11 do not use
the values of $V_r$ to measure the cluster velocity dispersion (which
is estimated from the line broadening of the combined spectra), this
analysis suggests that the spectra of the spaxels close to the
brightest stars are still significantly contaminated and provide an
artificial broadening of the combined lines. On the other hand, our
estimate is directly derived from the dispersion of the radial
velocities measured for each individual star in the field of view and
the statistics (52 stars) is quite good. Hence we do not see what kind
of mistake we could have committed to artificially decrease the value
of $\sigp(r)$.  We finally note that the center adopted by L11 is
slightly offset from ours (see Fig. \ref{vmap}).\footnote{Since there
  is a typo in the coordinates of the cluster centre quoted in
  equations (2) and (3) of L11, we have used their Figure 6 to
  approximately locate their centre on the HST/ACS-HRC image. As shown
  in the left-hand panel of Fig. \ref{vmap}, our centre is located
  $0.55\arcsec$ North and $0.40\arcsec$ East from theirs (differently
  from what is stated in L11).}  However, even if we adopt their
center and their radial bins, we still find results that are totally
inconsistent with a steep central cusp: we measure $\sigp=14.5\kms$ in
the central bin ($r<0.9\arcsec$, 17 stars), $\sigp=12.4\kms$ in the
second annulus ($0.9\arcsec<r<1.9\arcsec$; 27 stars), and
$\sigp=13.1\kms$ if a single bin at $r<1.3\arcsec$ (44 stars) is
adopted.

The derived central velocity dispersion also disagrees with the value
of $18.9\pm 0.8\kms$ determined by \citet{illingw76} and reported by
\citet{pryormey93}. We believe that, even in this case, the difference
comes from the fact that such a value has been obtained from the line
broadening of integrated-light spectra, which is prone to shot noise
bias. However, we also note that value of $18.9\kms$ refers to the
inner $6\arcsec-12\arcsec$ from the centre \citet[see Sect.II
  in][]{illingw76}, while the effective velocity dispersion quoted by
L11 has been computed for $r\le40\arcsec$. This cast doubts upon the
claimed agreement between the value quoted by L11 and that of
\citet{illingw76}. In addition, the latter author attempted to correct
the observed velocity dispersion, thus to derive the central value;
the result is quoted in his Table 5: $19.6\pm 5.7\kms$. Hence, within
such a large uncertainty, both our measurement and that of L11 are
formally in agreement with the central velocity dispersion obtained by
\citet{illingw76}.

\subsection{Comparison with models}
\label{sec:cfr_mod}
In this section theoretical models are used to reproduce the
observations and derive some constraints on the presence of a central
IMBH in NGC 6388. To this end, we follow two different and
complementary approaches. First, starting from a family of
self-consistent models admitting a central IMBH, we select the one
that best reproduces the observed density and velocity dispersion
profiles. This provides us with the corresponding structural and
kinematical parameters in physical units, including the IMBH
mass. Second, we use the observed density profile and include a
variable central point mass, to solve the spherical Jeans equation
(this is what is done also in L11). We therefore obtain the
corresponding family of velocity dispersion profiles, which are then
compared to the observations to constrain the BH mass.  The first
approach is more realistic in terms of the stellar mass-to-light ratio
($M/L$), since it takes into account various populations of stars with
different masses and different radial distributions (as it is indeed
expected and observed in mass segregated GCs), while in most
applications of the Jeans approach a constant $M/L$ profile is assumed
for the computation of the velocity dispersion \citep[but see, e.g.,
  Williams et al. 2009 and][for Jeans models with varying
  $M/L$]{lutz12}. Moreover the self-consistent modeling has the
advantage that the models stem from a distribution function which is
known a priori. Hence, the model consistency (i.e., the non-negativity
of the distribution function in the phase-space) is under control and
guaranteed by construction. However, the validity of the results is
limited to the case in which the assumed models are the correct
representation of the true structure and dynamics of the cluster. The
second approach is more general, but the resulting models could be
non-physical \citep[e.g.][]{BT87}.  In all cases we assume spherical
symmetry and velocity isotropy. These are reasonable assumptions for
NGC 6388, which does not show significant ellipticity \citep{harris}
and is thought to be quite dynamically evolved \citep[][see also
  L11]{dyn_clock}.

\begin{figure}[!hp]
\begin{center}
\includegraphics[scale=0.8]{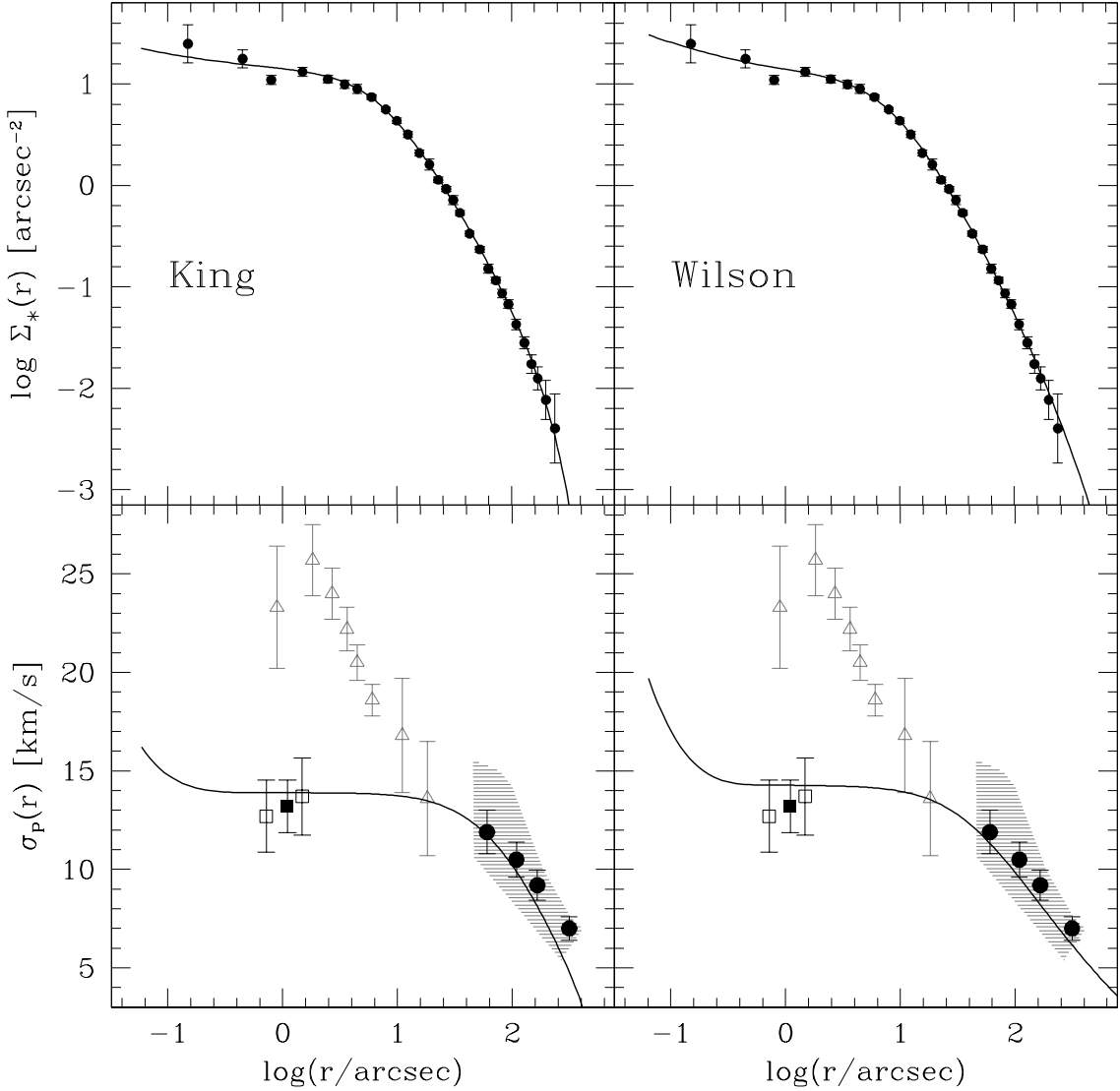}
\caption{Comparison between the observations and the best-fit
  self-consistent King and Wilson models with central IMBH (left- and
  right-hand panels, respectively). The observed density profile
  (solid circles in the upper panels) is from \citet{lanz07_imbh}. The
  observed velocity dispersion profile (symbols in the the lower
  panels) is the same as in Fig. \ref{vdisp}. Solid lines show the
  best-fit model results for the mass group corresponding to $0.8-0.9
  M_\odot$ stars. Model parameters are given in Table \ref{tab_mod}.}
\label{profs}
\end{center}
\end{figure}

\noindent
\emph {Self-consistent King and Wilson models $-$} In order to
reproduce the observed surface density and velocity dispersion
profiles we used self-consistent, multi-mass King (1966) and Wilson
(1975) models admitting the presence of a central IMBH
\citep{miocchi07}.\footnote{As detailed in \citet{miocchi07}, the IMBH
  is included following the \citet{BahcWolf76} phase-space
  distribution function. These models can be generated and freely
  downloaded from the Cosmic-Lab web site at the address: {\tt
    http://www.cosmic-lab.eu/Cosmic-Lab/Products.html}.}  We assumed
the same stellar mass spectrum as in \citet{lanz07_imbh}, consisting
of six mass bins, $0.1 M_\odot$ wide and ranging between $0.3 M_\odot$
and $0.9 M_\odot$, plus a seventh bin containing a population of $1.2
M_\odot$ white dwarfs, assumed to be the remnants of 4-8 $M_\odot$
stars. For the comparison with the observations we used only the
density and velocity dispersion profiles of the [0.8, 0.9] $M_\odot$
mass group, corresponding to turnoff and giant stars (the ones
effectively used in the observations).  By following the procedure
described in \citet{miocchi10}, we performed a parametric fit to the
observed density profile: this provided us with the model structural
parameters (gravitational potential, characteristic radius and ratio
between the BH mass and the cluster total mass).  Once the projected
density is fitted, the shape of the velocity dispersion profile is
univocally determined and we therefore applied only a vertical shift
(corresponding to a velocity scale factor) to adjust it to the
observations.  As shown in Figure \ref{profs}, both the King and the
Wilson models properly reproduce the observed profiles.
\footnote{Just for illustrative purposes, we show in Fig. \ref{vrr}
  the $\pm 3\sigma$ velocity dispersion profiles corresponding to the
  best-fit King and Wilson models (dotted and dashed curves,
  respectively). As apparent, only a few stars used in the computation
  of $\sigp(r)$ as described in Sect. \ref{sec:vdisp} are located
  below or above these lines, and the velocity dispersion profile that
  one would obtain by excluding them from the analysis (consistently
  with the assumption that the stellar system behaves as a perfect
  King or Wilson model) is still in agreement with the one quoted
  above.}  Once the distance of NGC 6388 is specified \citep[we
  adopted $d=13.2$ kpc from][]{ema08}, the assumed velocity scale
factor univocally determines the value of the cluster total mass
($M_{\rm tot}$).  In turn, given the BH-to-cluster mass ratio set by
the density best-fit, this provides us with the mass of the IMBH.  The
resulting values are listed in Table \ref{tab_mod}, with $M^{\rm
  sc}_{\rm BH}$ indicating the BH mass obtained from the
self-consistent modeling. This approach suggests an IMBH of $\sim 2000
M_\odot$.

\begin{figure}[!hp]
\begin{center}
\includegraphics[scale=0.8]{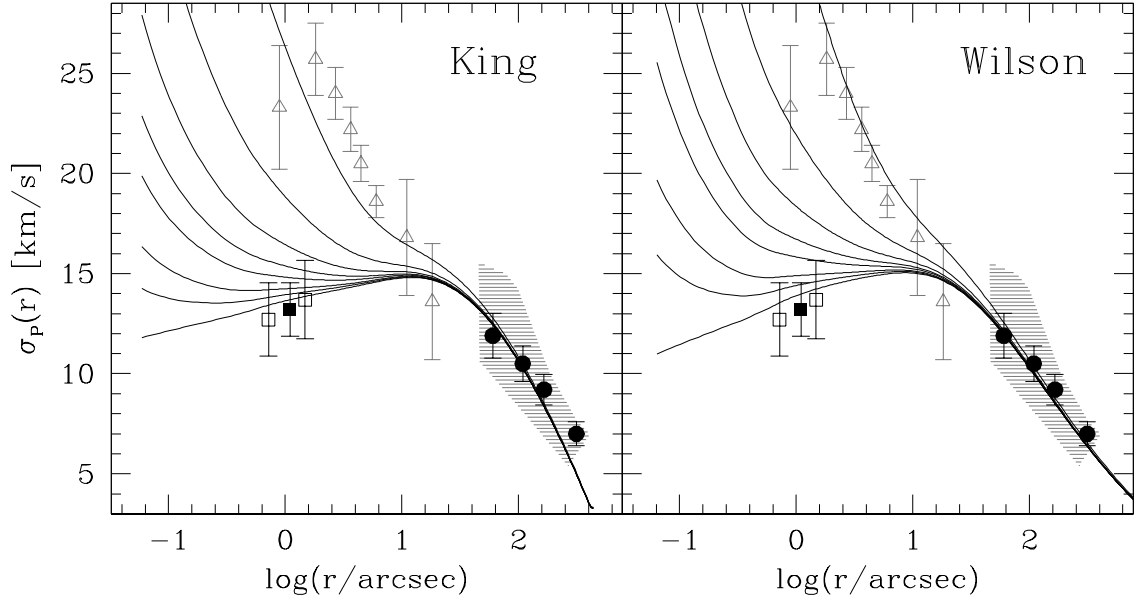}
\caption{Comparison of the observed velocity dispersion profile (the
  same as in Fig. \ref{vdisp}) with two families of Jeans models. {\it
    Left panel:} solid lines correspond to Jeans models calculated
  from the King density profile shown in the upper-left panel of
  Fig. \ref{profs} and by assuming different black hole masses: from
  bottom to top, $M^{\rm J}_{\rm BH}/M^{\rm sc}_{\rm BH}=0, 0.5, 1, 2,
  3, 5, 10, 30$, with $M^{\rm sc}_{\rm BH}=2147 M_\odot$. {\it Right
    panel:} the same as in the left panel, but for Jeans models
  calculated from the Wilson density profile shown in the upper-right
  panel of Fig. \ref{profs}. In this case, $M^{\rm sc}_{\rm BH}=2125
  M_\odot$.}
\label{jeans}
\end{center}
\end{figure}

\noindent
\emph{Jeans models $-$} Starting from the best-fit King and Wilson
density profiles (upper panels of Fig. \ref{profs}), we solved the
Jeans equation by admitting the presence of an IMBH with a mass
($M^{\rm J}_{\rm BH}$) ranging between zero and thirty times the
best-fit BH mass of the corresponding self-consistent model (i.e.,
$M^{\rm J}_{\rm BH}/M^{\rm sc}_{\rm BH}=0, ..., 30$, with $ M^{\rm
  sc}_{\rm BH}=2147 M_\odot$ and $2125 M_\odot$ for the King and
Wilson cases, respectively; see Table
\ref{tab_mod}). Results\footnote{For fixed BH mass the same velocity
  dispersion profile is expected in both approaches.  However, the
  comparison between $\sigp(r)$ shown in Fig. \ref{profs} and the
  curve corresponding to $M^{\rm J}_{\rm BH}/M^{\rm sc}_{\rm BH}=1$ in
  Fig. \ref{jeans} reveals a slightly different shape.  This is
  because the profile shown in Fig. \ref{jeans} corresponds to the
  velocity dispersion produced by all cluster stars, instead of that
  produced by the [0.8, 0.9] $M_\odot$ mass group only (which is shown
  in Fig. \ref{profs}). In fact, the adopted Jeans equation approach
  is unable to distinguish among different mass groups (or, in other
  words, it assumes a constant $M/L$ ratio).} are shown in Figure
\ref{jeans} and indicate that our observed velocity dispersion profile
is consistent with both the absence of a central IMBH, and the
presence of a black hole with a mass up to $\sim 2000
M_\odot$. Instead we find that in order to match the L11 profile an
IMBH of $\approx 6\times 10^4 M_\odot$ should be assumed, in rough
agreement with the value quoted by these authors.

The results of our analysis are consistent with both no IMBH and a
black hole of no more than $2000 M_\odot$ in the centre of NGC 6388.
In particular, its presence is mainly constrained by the cusp of the
density profile, while its mass is set by the observed velocity
dispersion.\footnote{Indeed, the difference in the derived value of
  $M_{\rm BH}$ with respect to our previous estimate ($5700 M_\odot$)
  is mostly due to the fact that in \citet{lanz07_imbh} we used a
  photometric estimate of the cluster total mass under the common
  assumption of a stellar mass-to-light ratio $M/L_V=3$
  \citep[e.g.][]{djorg93}. Here, instead, the measured velocity
  dispersion provides us with a dynamical estimate of the cluster
  total mass, which turns out to be a factor of $\sim 2.2$
  smaller. This is in agreement with recent results suggesting that
  $M/L_V\sim 1.3$ in GCs \citep[see, e.g.][]{sollima12}.}  In fact,
the steep central increase expected in $\sigp(r)$ is not visible in
our observations, because, for a $2000 M_\odot$ BH, it is predicted to
occur at a very small distance from the centre ($r\lsim 0.5\arcsec
=0.03$ pc), where we measured only 9 stars.  Note that such a small
cusp radius is in agreement with the standard computation of the BH
sphere of influence ($r_{\rm BH} \sim G M_{\rm BH}/\sigma^2$, with
$\sigma$ being the unperturbed velocity dispersion): in fact, assuming
$\sigma\sim14\kms$, one gets $r_{\rm BH}\sim 0.7\arcsec$.  We finally
conclude by noticing that the discussed family of self-consistent
models are incompatible with the velocity dispersion profile derived
by L11 (triangles in Fig.  \ref{profs}). In fact, in these models the
velocity dispersion cusp due to the presence of an IMBH always appears
within the innermost $3\arcsec-5\arcsec$ from the centre and the main
effect of increasing the BH mass is to make the cusp steeper, while
there is no mean of moving it outward, at $10\arcsec$ or more, where
L11 already observed a significant increase of $\sigp(r)$.

\section{SUMMARY AND CONCLUSIONS}
\label{sec:concl}
By using the high spatial resolution spectrograph SINFONI, and the
multi-object spectrograph FLAMES/MEDUSA installed at the ESO VLT, we
measured the radial velocities and metallicities of $\sim 600$ stars
in the direction of the Galactic globular cluster NGC 6388.  Based on
the measurements of $V_r$ and [Fe/H] obtained for the FLAMES sample,
we determined the systemic radial velocity and mean metallicity of NGC
6388: $V_{\rm sys}=82.0\pm 0.5 \kms$ and $\langle$[Fe/H]$\rangle=
-0.40\pm 0.08$ (which are in good agreement with the values quoted in
the Literature). We also set the criteria for cluster membership:
$60\le V_r\le 105\kms$ and $-0.64<$[Fe/H]$<-0.16$. This provided us
with 52 members located at $0.2<r<1.9\arcsec$ and 276 between
$18\arcsec$ and $600\arcsec$ from the SINFONI and the FLAMES samples,
respectively.

The radial velocities of cluster members were used to study the
systemic rotation and velocity dispersion of NGC 6388.  We found
evidence of systemic rotation with amplitude $A_{\rm rot}\sim 8 \kms$
in the innermost $2\arcsec$ (0.13 pc), decreasing to $A_{\rm rot}=3.2
\kms$ at $18\arcsec<r<160\arcsec$.  In the outermost regions no
rotation signature is detected, most probably because the sample is
not rich enough and the (largely dominant) velocity dispersion hides
the rotation signal.  The cluster velocity dispersion profile, both in
the very center ($0.2\arcsec<r<1.9\arcsec$) and in the external
regions (out to $\sim 600\arcsec$) has been derived from the
dispersion of {\it individual star} radial velocities. Results
indicate a central value of $\sim 13-14\kms$, a flat behavior out to
$\sim 60\arcsec$ and a decreasing trend outwards. No evidence of a
central cusp in $\sigma_P(r)$ is detected in our data.

These results desagree with those published by L11, who find a steep
cusp at $r<10\arcsec$ and a central velocity dispersion of $\sim
23-25\kms$.  The reason for such a discrepancy is not completely
clear, but it is likely related to the fact that L11 measure $\sigp$
from the line broadening of integrated light spectra. As it is widely
discussed in the literature \citep[e.g.,][]{dubath97,jay10}, this
procedure is prone to strong biases in the case of Galactic GCs. In
fact the acquired spectra can be dominated by the contribution of a
few bright stars, instead of effectively sampling the underling
stellar distribution. Certainly, using high-resolution IFU
spectroscopy, applying procedures to correct for the shot noise
induced by giants, and measuring the line broadening from the combined
spectra of several spaxels in the same radial annulus (as it is done
in L11) may alleviate the problem.  However, $(i)$ the fact that it is
hard to imagine methodological errors able to artificially lower the
value of $\sigma_P(r)$ if estimated from individual radial velocities
(at least if the statistics is good, as it is the case in our study
which comprises 52 stars), $(ii)$ the apparently high degree of color
correlation among several spaxels close to the brightest stars in the
$V_r$ map of L11, and $(iii)$ the disagreement (in terms of both
velocity dispersion and radial velocity map; see Figs. \ref{vdisp} and
\ref{vmap}) between the two methodologies suggest that measuring
$\sigp$ from the integrated light spectra can be fallacious.  Indeed,
similar discrepancies between the "integrated light" and the
"statistical" approaches are found also in the case of $\omega$
Centauri, where the latter method is applied to proper motion
measurements \citep[see][]{jay10,noyola10}.

The derived velocity dispersion profile is consistent with the
presence of an IMBH at most as massive as $\sim 2000 M_\odot$.  Given
the measured central velocity dispersion (13-14 $\kms$), the derived
IMBH mass is in agreement with the expectations of the $M_{\rm
  BH}-\sigma$ relation within its quoted uncertainties \citep[see
  e.g.,][]{gultek09}. More sophisticated models are needed to put
stronger constraints.  However, we believe that to obtain a firm
answer it is first necessary to precisely understand why integrated
light spectra and the measurement of individual star radial velocities
(or proper motions) yield to different values of the velocity
dispersion. It is also necessary to collect more data, providing the
full velocity profile \citep[rotation, velocity dispersion, and also
  the higher order terms; see, e.g., the discussion in][]{gebh05}. To
this end, high spatial resolution photometry and spectroscopy are key
observational tools, and new sophisticated methods for the analys of
integral field unit data cubes \citep[e.g.,][]{kamann13} are very
promising.  Meanwhile the James Webb Space Telescope and the planned
ground-based extremely large telescopes will become available, for a
significant progresses in this field of the research it is certainly
worth to push even further the usage of the HST and the current
generation of spectrometers at 8-10m class telescopes, with integral
field capabilities and with adaptive optics.

\acknowledgements We warmly thank the referee for the careful reading
of the manuscript and the useful suggestions that helped improving the
quality of the paper. This research is part of the project COSMIC-LAB
funded by the European Research Council (under contract
ERC-2010-AdG-267675).

\appendix
\section{IMPACT OF UNRESOLVED BACKGROUND AND BLENDS ON THE DETERMINATION OF $V_r$}
\label{app}
\begin{figure}[!hp]
\begin{center}
\includegraphics[scale=0.7]{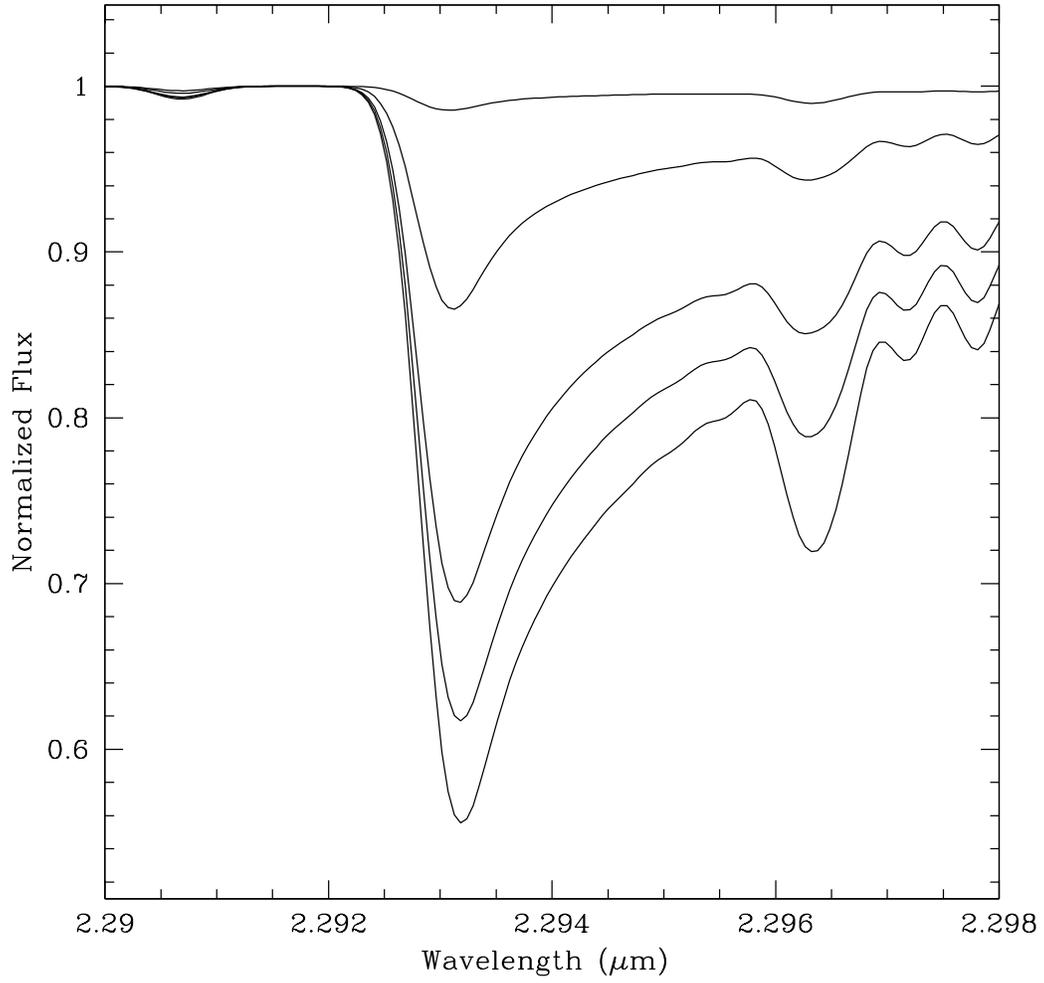}
\caption{Synthetic spectra in the main CO band-head region plotted for
  five different values of the effective temperature (from bottom to
  top: $T_{\rm eff}=3500, 4000, 4500, 5000, 5500$ K).}
\label{fig_app}
\end{center}
\end{figure}

We estimated that the contamination from unresolved background stars
(that, in principle, could artificially lower the derived velocity
dispersion) is totally negligible. In fact, the dominant contribution
to the background is due to unresolved $I\gsim 19$ stars which have
effective temperatures $T_{\rm eff} \gsim 5500$ K. On the other hand,
for $T_{\rm eff}>4500$ K the depth of the CO band-heads rapidly
decreases for increasing temperature (see Figure \ref{fig_app}), and
at $T_{\rm eff}=5500$ K it is totally negligible with respect to the
CO depth at 5000 K (the temperature of our faintest target).  We
therefore conclude that the unresolved background does not affect the
determination of $V_r$ for our targets. We also stress that, even if,
we took into account only the brightest targets (45 stars with
$I<16.5$), the resulting central velocity dispersion would remain
essentially unchanged: $\sigp=13.8\pm1.5\kms$. 

In the case of non-isolated targets, we estimated the effect of blends
on the computation of $V_r$ as follows \citep[note that more complete
  methods have been proposed in the literature, but they are far too
  much sophisticated for the quality of our data; see,
  e.g.,][]{kamann13}. By convolving each star with its point spread
function, we estimated the fraction of light from neighbouring objects
that contaminates the spaxel of the candidate SINFONI target. We found
that the contribution of stars detected in the HST image but not
corresponding to a SINFONI centroid was always negligible ($<5\%$).
This is not surprising, since these stars are fainter by at least
1-1.5 magnitudes than the SINFONI targets.  Instead, we found seven
couples of SINFONI targets where the spaxel corresponding to the
fainter star (and used to measure its $V_r$; see Sect. \ref{sec:vr})
was contaminated by 20-25\% from the light of the neighbouring
(brighter) star. For each of these couples, we simulated the spectrum
of the fainter object blended with that of the neighbouring star, by
taking into account the percentage of contamination and the
differences in flux, temperature and radial velocity between the two
``companions''. We then measured the radial velocity from this blended
spectrum and we compared it to the input one (i.e., the value
previously measured from the observed spectrum of the fainter
target). In four cases, the resulting value of $V_r$ was in agreement
with the input one within the uncertainties. Discrepancies of about
$-5 \kms$ have been found only for the remaining three cases for which
the contaminating (bright) star had $V_r$ smaller than the
contaminated (fainter) one. Such a behaviour is due to the asymmetry
of the CO band-heads: when the radial velocity of the contaminating
star is larger than that of the fainter target, the offset is hidden
within the CO band shape, otherwise it is detectable as a shift in
$V_r$ towards lower values. We therefore conclude \citep[in agreement
  with the findings of][]{kamann13} that blending is not a major issue
for the measured radial velocities but, to be conservative and to
avoid any possible concern about the velocity measurements, we decided
to remove the three most affected targets from the sample. Note that
including or excluding these stars (or even all the seven targets with
larger contamination) has no impact on the derived velocity
dispersion.

% *********************************** TABLES **********************************************
\begin{table}
\begin{center}
\begin{tabular}{rccccccccc}
\hline \hline
id  & RA & DEC & $B$ & $V$ & $I$ & $\widetilde V_r$& $e_{\widetilde Vr}$ &  [Fe/H] & data set\\
\hline
10000131 & 264.0715464 & -44.7349731 &    -- & 14.94 & 12.82 &   23.5 & 2.3 &    -- & 1 \\
10000216 & 264.0720428 & -44.7353238 &    -- & 15.12 & 12.77 &  -18.1 & 2.0 &    -- & 1 \\
10000027 & 264.0712349 & -44.7352065 &    -- & 15.53 & 13.80 &   13.5 & 2.0 &    -- & 1 \\
10000325 & 264.0724705 & -44.7351295 &    -- & 15.53 & 13.78 &  -18.7 & 2.7 &    -- & 1 \\
... & & & & & & & & & \\
   30828 & 264.1096197 & -44.7367217 & 16.90 & 14.96 &   --  &   -4.8 & 0.3 & -0.49 & 2 \\
 7000555 & 264.0853010 & -44.7688900 & 16.90 & 15.27 & 13.40 &  -13.8 & 0.2 & -0.48 & 2 \\
 7001285 & 264.0194000 & -44.7440720 & 17.34 & 15.98 & 14.37 &    1.1 & 0.1 & -0.39 & 2 \\
 7001303 & 264.1128920 & -44.7231290 & 17.74 & 16.32 & 14.61 &   10.0 & 0.2 & -0.41 & 2 \\
\hline
\end{tabular}
\caption{Identification number (id), coordinates (in degree),
  magnitudes in the $B$, $V$, and $I$ bands, radial velocity and its
  uncertainty in $\kms$ ($\widetilde V_r$ and $e_{\widetilde Vr}$, respectively),
  [Fe/H] abundance, and flag for the data set (fl=1 meaning SINFONI,
  fl=2 indicating FLAMES) for all the stars selected as cluster
  members and used to compute the rotational velocity and velocity
  dispersion of NGC 6388. No values of metallicity are given for the
  SINFONI targets because of their large uncertainty. The complete
  version of the table is available online.}
\label{tab_vr}
\end{center}
\end{table}

\begin{table}
\begin{center}
\begin{tabular}{rccccc}
\hline \hline
$r_i$ & $r_e$  & $r_m$ & $N_\star$ & $\sigma_P$ & $e_{\sigma P}$ \\
\hline
   0.2 &   1.9 &   1.10 & 52 & 13.20 & 1.33\\
  18.5 &  86.0 &  60.17 & 59 & 11.90 & 1.11\\
  86.0 & 135.0 & 108.79 & 70 & 10.50 & 0.89\\
 135.0 & 205.0 & 165.23 & 72 &  9.20 & 0.77\\
 205.0 & 609.0 & 312.75 & 71 &  7.00 & 0.59\\
\hline
\end{tabular}
\caption{Velocity dispersion profile of NGC 6388. The three first
  columns give the internal, external and mean radii (in arcseconds) of
  each considered radial bin ($r_m$ is computed as the average
  distance from the centre of all the stars belonging to the bin),
  $N_\star$ is the number of star in the bin, $\sigma_P$ and
  $e_{\sigma P}$ are the velocity dispersion and its rms error (in
  $\kms$), respectively.}
\label{tab_vdisp}
\end{center}
\end{table}

\begin{table}
\begin{center}
\begin{tabular}{rcccc}
\hline \hline
Model  & $W_0$& $r_c$& $M_{\rm tot}$ & $M^{\rm sc}_{\rm BH}$ \\ 
\hline
King   &  11  & 6.4  & $1.2 \times 10^6$ & 2147 \\
Wilson & 11   & 6.8  & $1.2 \times 10^6$ & 2125 \\  
\hline
\end{tabular}
\caption{Parameters of the King and Wilson models that best fit the
  observed surface density and velocity dispersion profiles (see
  Figure \ref{profs}). $W_0$ is a dimensionless parameter proportional
  to the gravitational potential at the cluster center, $r_c$ is the
  core radius in arcseconds (defined as the radius at which the
  projected stellar density drops to half its central value), $M_{\rm
    tot}$ and $M^{\rm sc}_{\rm BH}$ (in units of $M_\odot$) are,
  respectively, the total cluster mass and the mass of the central
  IMBH estimated from the self-consistent modelling approach. }
\label{tab_mod}
\end{center}
\end{table}

\newpage
% ******************************* FIGURES **********************************************

\end{document}